\documentclass[12pt]{article}
\usepackage{natbib}
\usepackage{url}
\usepackage[colorlinks,citecolor=blue,urlcolor=blue]{hyperref}
\usepackage{booktabs}
\usepackage{amsfonts}
\usepackage{nicefrac}
\usepackage{xcolor}
\usepackage{bm}
\usepackage{amsthm,amsmath,amsopn,amssymb}
\usepackage{verbatim}
\usepackage[ruled,linesnumbered]{algorithm2e}
\usepackage{graphicx}
\usepackage{threeparttable}
\usepackage{multicol, multirow}
\usepackage{enumitem}
\usepackage{wrapfig}
\usepackage{caption}
\usepackage{subcaption}  
\usepackage{accents}
\usepackage[noend]{algpseudocode}
\defcitealias{li2024a}{Li et al. (2024a)}
\defcitealias{li2024b}{Li et al. (2024b)}
\newcommand{\blind}{0}

\newtheorem{theorem}{Theorem}

\newtheorem{proposition}{Proposition}
\theoremstyle{definition}

\makeatletter
\def\widebar{\accentset{{\cc@style\underline{\mskip10mu}}}}
\makeatother

                \def\y{\bm{y}}

\def\RR{\mathbb{R}}

\usepackage[left=2.5cm,right=2.5cm,top=2.5cm,bottom=2.5cm]{geometry}

\def\blind{0}
\begin{document}

\def\spacingset#1{\renewcommand{\baselinestretch}%
{#1}\small\normalsize} \spacingset{1}


\if0\blind
{
\title{\bf Core-elements Subsampling for Alternating Least Squares
}
\author{Dunyao Xue\\
        Institute of Statistics and Big Data, \\
        Renmin University of China, Beijing, China \\
       Mengyu Li\thanks{Corresponding author, mengyuli@tsinghua.edu.cn}\\
       Department of Statistics and Data Science, Tsinghua University, Beijing, China \\
       Jingyi Zhang
       \\
       School of Science, Department of Mathematics\\
       Beijing University of Posts and Telecommunications, 
       Beijing, China\\
       Cheng Meng
       \\
       Center for Applied Statistics, Institute of Statistics and Big Data, \\
       Renmin University of China, Beijing, China. \\
}
\date{}
  \maketitle
} \fi

\if1\blind
{
  \bigskip
  \bigskip
  \bigskip
  \begin{center}
    {\LARGE\bf Core-elements Subsampling for Alternating Least Squares}
\end{center}
  \medskip
} \fi

\bigskip

\begin{abstract}
In this paper, we propose a novel element-wise subset selection method for the alternating least squares (ALS) algorithm, focusing on low-rank matrix factorization involving matrices with missing values, as commonly encountered in recommender systems. While ALS is widely used for providing personalized recommendations based on user-item interaction data, its high computational cost, stemming from repeated regression operations, poses significant challenges for large-scale datasets.
To enhance the efficiency of ALS, we propose a core-elements subsampling method that selects a representative subset of data and leverages sparse matrix operations to approximate ALS estimations efficiently.
We establish theoretical guarantees for the approximation and convergence of the proposed approach, showing that it achieves similar accuracy with significantly reduced computational time compared to full-data ALS. Extensive simulations and real-world applications demonstrate the effectiveness of our method in various scenarios, emphasizing its potential in large-scale recommendation systems.

\end{abstract}

\noindent%

\vfill

\newpage
\spacingset{1.75} 

\section{Introduction}
\label{sec:intro}

Recommender systems are designed to provide personalized suggestions to users by modeling their preferences over items \citep{lu2012recommender,bi2017group,RecommenderSystemsReview}. These systems utilize explicit feedback, such as user ratings, and implicit feedback, including user activities such as purchasing, viewing, or searching for items.

Matrix factorization (MF) techniques have become fundamental in developing effective recommender systems \citep{lee2000algorithms,sun2016guaranteed}. Among these, alternating least squares (ALS) \citep{WOS:000256870200030,jain2013low,takacs2012alternating} is a robust MF algorithm capable of handling both explicit and implicit feedback data. Unlike the classical singular value decomposition (SVD) \citep{klema1980singular,abdi2007singular,zhao2024distributed}, ALS efficiently manages missing values in the user-item matrix, thereby improving its applicability in a broader range of real-world scenarios.

Despite its effectiveness, low-rank matrix factorization using alternating least squares can be computationally intensive due to the numerous regression operations required, particularly when dealing with large initial matrices. For instance, given an initial matrix $\boldsymbol{R}\in \RR^{n_u \times n_m}$, the time complexity of ALS algorithms is $O\left(n_f^2\left(\text{nnz}(\boldsymbol{R}) + n_f n_u + n_f n_m\right) n_t\right)$ \citep{WOS:000256870200030}, where $n_f$ denotes the rank of the decomposed low-rank matrix, $n_t$ represents the number of iterations, and $\text{nnz}(\cdot)$ denotes the number of non-zero elements of a matrix.

An effective strategy to improve computational efficiency is to estimate the model on a subset of observations, commonly referred to as the coresets approach, subsampling, or subset selection. By strategically selecting representative observations using methods such as leverage scores \citep{ma2015leveraging, ma2014statistical, han2023leverage, zhong2023model, shimizu2023improved}, influence functions \citep{ting2018optimal}, predictive inference-based subsampling \citep{wu2024optimal}, various optimality criteria \citep{wang2018optimal, yu2023information, wang2019information}, and other probabilistic or deterministic techniques, coresets substantially reduce computational burdens while retaining most of the information in large datasets. These approaches have proven effective across a range of statistical learning tasks, including least squares regression \citep{ ma2014statistical, ma2015leveraging, ting2018optimal, meng2017effective, drineas2006sampling, wang2019information, wang2021orthogonal, ma2022asymptotic}, generalized linear models \citep{wang2018optimal, ai2021optimal, yu2022optimal, yu2024subsampling}, nonparametric regression \citep{ma2015efficient, meng2020more, meng2022smoothing, zhang2024independence}, quantile regression \citep{wang2021optimal, ai2021quantile}, model-free methods \citep{meng2021lowcon, yi2023model}, machine learning \citep{wang2018optimal, han2023leverage}, time series analysis \citep{xie2019online, xie2023optimal}, and optimal transport \citep{li2023efficient, hu2024sampling, li2023efficient}. Beyond the computational benefits, coresets play a pivotal role in measurement-constrained problems \citep{zhang2023optimal, meng2021lowcon} and privacy-preserving contexts \citep{wang2019information, balle2020privacy}. For a comprehensive overview, we refer readers to \citet{li2020modern} and \citet{yu2024review}.

There have been several efforts to accelerate the alternating least squares algorithm through subsampling techniques. \citet{WOS:000264173600051} introduced a sampling ALS ensemble method that repeatedly samples elements from the rating matrix according to a predefined probability distribution, applies ALS to each sampled matrix, and then aggregates the results through weighted averaging. Furthermore, \citet{WOS:000458973703072} used a row sampling strategy based on leverage scores \citep{han2023leverage,  ma2014statistical} to improve the efficiency of ALS for tensor data. However, these sampling approaches exhibit some limitations. Element-wise sampling directly from the rating matrix can result in substantial information loss, and leverage score-based sampling incurs high computational costs due to the need to compute leverage scores. Although approximations of sampling probabilities can alleviate some of this computational burden, they often compromise accuracy. Therefore, we aim to develop a novel sampling method that significantly accelerates the traditional ALS algorithm while preserving high accuracy.

Beyond sampling techniques, various methods have been proposed to accelerate ALS. For example, \citet{WOS:000256870200030} used parallel computing to reduce the computational costs associated with repeated regression operations. \citet{pilaszy2010fast} applied the Sherman–Morrison formula (SMF) to accelerate computations. Additionally, \citet{hastie2015matrix} explored the relationship between matrix completion and singular value decomposition, proposing a new iteration formulation for ALS. However, these approaches are beyond the scope of this paper, as sampling algorithms do not modify the core structure of the ALS algorithm and can be integrated with other acceleration techniques. Consequently, our focus is on developing advanced sampling algorithms within the fundamental ALS framework, rather than introducing entirely new algorithmic formulations.

\begin{figure}[h]
    \includegraphics[height=0.2\textheight]{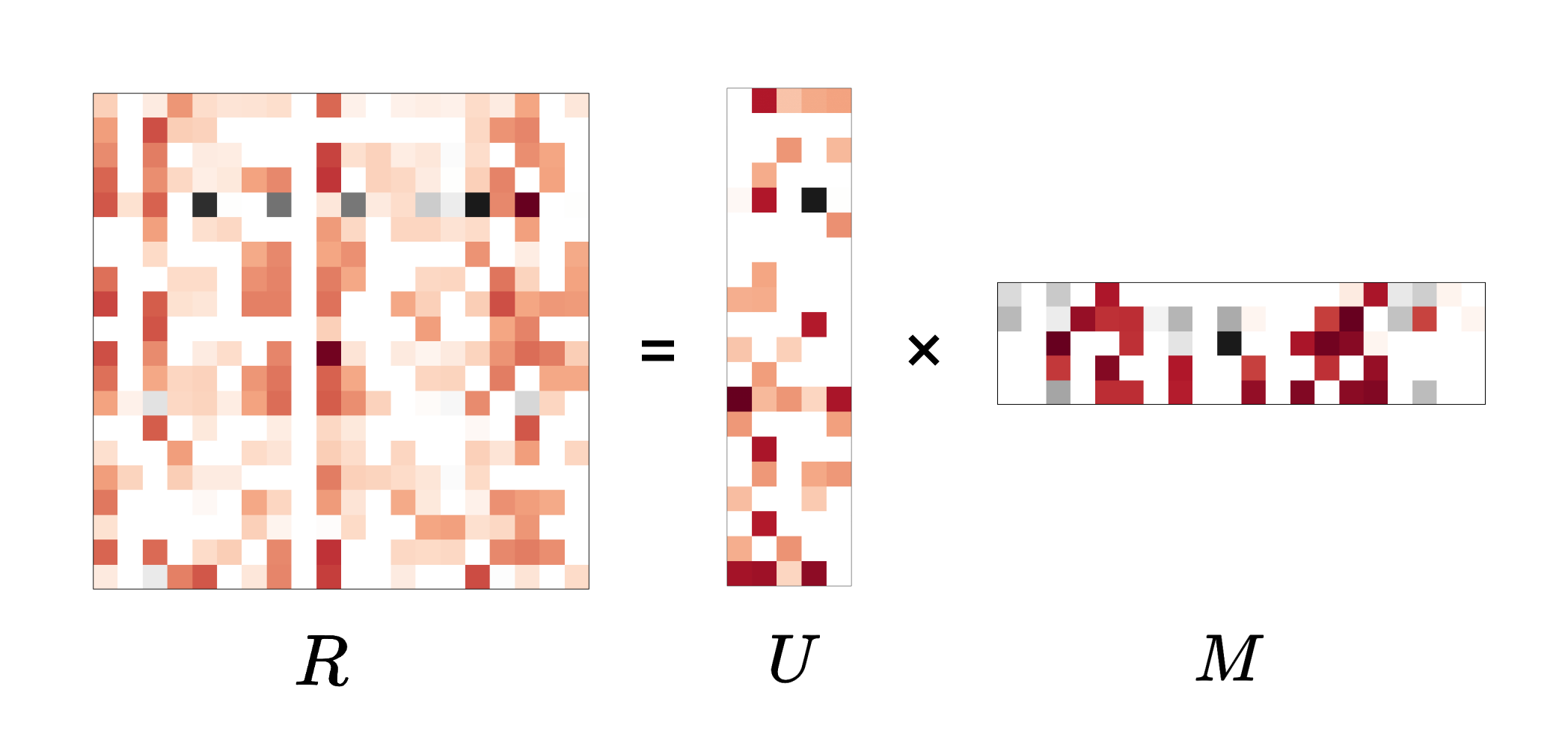}
    \centering
    \caption{Illustration of matrix factorization. Values less than $5\%$ of the maximum value are displayed as white, showing that both $\boldsymbol{U}$ and $\boldsymbol{M}$ are numerically sparse. }
    \label{sparse matrix}
\end{figure}

In recommender systems, we observe that both non-negative matrix factorization and matrix factorization with regularization terms tend to produce low-rank matrices with a large number of small values, resulting in numerical sparsity, as illustrated in Fig.~\ref{sparse matrix}. 
Sampling rows from such numerically sparse matrices, $\boldsymbol{U}$ or $\boldsymbol{M}$, can be significantly inefficient because most sampled elements are not informative. In contrast, the core-elements approach \citep{li2024a}, an element-wise subset selection method specifically designed for sparse matrices in linear and nonparametric regressions, effectively addresses this limitation by preserving the most informative components in the data.
Our extensive experimental results indicate that ALS often exhibits similar numerical sparsity characteristics as shown in Fig.~\ref{sparse matrix}. Therefore, extending the core-elements method to the ALS framework can be an effective solution to address this challenge.

Building on the foundations laid by these previous studies, in this paper we propose an efficient and scalable approach for approximating alternating least squares estimation in matrix factorization.

\textbf{Major contributions.} We summarize our contributions as follows.

\textit{First, we propose a novel core-element sampling framework Core-ALS for alternating least squares that selects important entries from the entire matrix and couples this with sparse matrix operations at each regression step.} This formulation efficiently approximates penalized least squares estimation and fully uses the information in the rating matrix $\boldsymbol{R}$. 
\textit{Second, we establish theoretical guarantees for the iterative setting.}
We derive per-iteration $(1+\epsilon)$ approximation bounds for the sketched regressions and show convergence of the Core-ALS method under mild conditions. Time complexity reductions specific to alternating updates and sparse matrix multiplications are also provided.
\textit{Third, we further improve the computational efficiency of \citetalias{li2024a} and  \citetalias{li2024b}.} We integrate partial quicksort \citep{martinez2004partial} into our \textit{Core Sparse Matrix Multiplication} so that thresholding-and-sampling is performed on the fly, which avoids full sorting and yields substantial computational speedups. Moreover, we develop a fast variant of Core-ALS that substantially lowers the computational cost.
\textit{Fourth, we show superior accuracy and efficiency from experiments.} 
Extensive simulations confirm the method’s effectiveness and efficiency for both model fitting and prediction, with particularly strong results on classical recommendation metrics including NDCG@k and Hit@k.

The remainder of this paper is organized as follows. In Section~\ref{ch:2}, we introduce the alternating least squares algorithm and core-elements for penalized regression splines. Section~\ref{ch:3} develops the core-elements algorithm for ALS, and Section~\ref{ch:4} discusses its theoretical properties. We then assess the performance of the proposed estimator through extensive experiments on synthetic and real-world data in Sections~\ref{ch:5} and \ref{ch:6}, respectively. Additional details omitted from the main text and technical proofs can be found in the Supplementary Material. All R code used to reproduce the results in this paper is available at
\url{https://github.com/sapphirexdy}.

\section{Background}\label{sec:background} 

Here we summarize the notation used throughout the paper. Specifically, matrices are represented by uppercase boldface italic letters, such as $\boldsymbol{X}$, while vectors are denoted by lowercase boldface italics letters such as $\boldsymbol{x}$. Scalars are represented using regular, non-bold typefaces. For any vector $\boldsymbol{x}$, we denote its $\ell_p$ norm by $\|\boldsymbol{x}\|_p$, and its Euclidean norm (i.e., the $\ell_2$ norm) is abbreviated as $\|\boldsymbol{x}\|$. When referring to matrices, the spectral norm is represented by $\|\boldsymbol{X}\|_2$, and the Frobenius norm is denoted as $\|\boldsymbol{X}\|_F$.

\label{ch:2}
\subsection{Alternating Least Squares Algorithm}

Consider a sparse\footnote{Here, the term ``sparse'' does not refer to a matrix with a large number of zero elements, but rather to one with a large proportion of missing values.} rating matrix $\boldsymbol{R}=(r_{i j}) \in \mathbb{R}^{n_u \times n_m}$ representing $n_u$ users and $n_m$ items, and a chosen implicit vector dimension $n_f$. The objective is to find two factor matrices $\boldsymbol{U} \in \mathbb{R}^{n_u \times n_f}$ and $\boldsymbol{M} \in \mathbb{R}^{n_m \times n_f}$ such that the relationship 
$
\boldsymbol{R} = \boldsymbol{U}\boldsymbol{M}^{\top}
$
holds in the absence of missing values.

More specifically, let $\boldsymbol{U}=(\boldsymbol{u}_i)$, where $\boldsymbol{u}_i \in \mathbb{R}^{1 \times n_f}~(i=1, \ldots, n_u)$ denotes the $i$th row of $\boldsymbol{U}$, and let $\boldsymbol{M}=(\boldsymbol{m}_j)$, where
$\boldsymbol{m}_j \in \mathbb{R}^{1\times n_f}~(j=1, \ldots, n_m)$ is the $j$th row of $\boldsymbol{M}$. If the non-missing values of $\boldsymbol{R}$ are fully predictable and $n_f$ is sufficiently large, we could expect that $$r_{i j}=\boldsymbol{u}_i \boldsymbol{m}_j^{\top} \quad \forall (i, j)\in [n_u] \times [n_m].$$
In practice, to estimate $\boldsymbol{U}$ and $\boldsymbol{M}$, the alternating least squares algorithm minimizes the empirical loss:
\begin{equation*} 
\mathcal{L}^{e m p}(\boldsymbol{R}, \boldsymbol{U}, \boldsymbol{M})=\frac{1}{n} \sum_{(i, j) \in I} (r_{i j}-\boldsymbol{u}_i \boldsymbol{m}_j^{\top})^2,
\end{equation*}
where $I$ is the index set corresponding to the non-missing values of $\boldsymbol{R}$ and $n = |I|$.

To prevent overfitting, a common approach is to append a Tikhonov regularization term \citep{tikhonov1977solutions} to the empirical risk:
\begin{equation}\label{eq:loss-reg}
\mathcal{L}_\lambda^{r e g}(\boldsymbol{R}, \boldsymbol{U}, \boldsymbol{M})=\mathcal{L}^{e m p}(\boldsymbol{R}, \boldsymbol{U}, \boldsymbol{M})+\lambda\left(\|\boldsymbol{U} \boldsymbol{\Gamma_U}\|^2+\|\boldsymbol{M}\boldsymbol{\Gamma_M} \|^2\right),
\end{equation}
for certain suitably selected Tikhonov matrices $\boldsymbol{\Gamma_U}$  and  $\boldsymbol{\Gamma_M}$.

Among the various types of regularization terms, the weighted $\lambda$-regularization works well empirically, preventing overfitting even as the number of features, $n_f$, or the number of ALS iterations increases \citep{WOS:000256870200030}.
Therefore, we focus on the regularized formulation \eqref{eq:loss-reg} of ALS. In the rest of this paper, ALS refers to the following objective function:
\begin{equation}
    f(\boldsymbol{U}, \boldsymbol{M})=\sum_{(i, j) \in I}\left(r_{i j}-\boldsymbol{u}_i \boldsymbol{m}_j^{\top}\right)^2+\lambda\Big(\sum_i n_{u_i}\|\boldsymbol{u}_i\|^2+\sum_j n_{m_j}\|\boldsymbol{m}_j\|^2\Big),
    \label{object function}
\end{equation}
where $n_{u_i}$ and $n_{m_j}$ denote the number of ratings of user $i$ and item $j$, respectively. This corresponds to taking $\boldsymbol{\Gamma_U}=\operatorname{diag}\left(\sqrt{n_{u_i}}\right)$ and $\boldsymbol{\Gamma_M}=\operatorname{diag}\left(\sqrt{n_{m_j}}\right)$ in \eqref{eq:loss-reg}. 

Let $f(\boldsymbol{U}, \boldsymbol{M})$ take the partial derivatives for $\boldsymbol{U}$ and $\boldsymbol{M}$, respectively, the specific iteration format of alternating least squares algorithm can be obtained as follows:

Step 1 (Initialization). Initialize matrix $\boldsymbol{M}$ randomly and let $\boldsymbol{\widehat{M}} = \boldsymbol{M}$.

Step 2 (Update $\boldsymbol{U}$). Fix $\boldsymbol{\widehat{M}}$ and solve for $\boldsymbol{U}$ by \eqref{U format}:
\begin{equation}    \widehat{\boldsymbol{u}}_i=\Big(\boldsymbol{\widehat{M}}_{I_i^U}^{\top} \boldsymbol{\widehat{M}}_{I_i^U}+\lambda n_{u_i} \boldsymbol{E}\Big)^{-1} \boldsymbol{\widehat{M}_{I_i^U}^{\top}} \boldsymbol{R}^{\top}\big(i, I_i^U\big), \quad \forall i= 1,\dots,n_u,
\label{U format}
\end{equation}
where $I_i^U$ denotes the set of indices of non-missing values in the $i$th row of $\boldsymbol{R}$, $\boldsymbol{\widehat{M}}_{I_i^U}$ denotes the sub-matrix of $\boldsymbol{\widehat{M}}$ where rows $r \in I_i^U$ are selected, and $\boldsymbol{R}\left(i, I_i^U\right)$ is the row vector where columns $j \in I_i^U$ of the $i$th row of $\boldsymbol{R}$ are selected. $\boldsymbol{E}$ is the $n_f\times n_f$ identity matrix.

Step 3 (Update $\boldsymbol{M}$). Fix $\boldsymbol{\widehat{U}}$ and solve for $\boldsymbol{M}$ by \eqref{M format}:
\begin{equation}
\widehat{\boldsymbol{m}}_j=\Big(\boldsymbol{\widehat{U}}_{I_j^M}^{\top} \boldsymbol{\widehat{U}}_{I_j^M}+\lambda n_{m_j} \boldsymbol{E}\Big)^{-1} \boldsymbol{\widehat{U}_{I_j^M}^{\top}} \boldsymbol{R}\big(I_j^M, j\big), \quad \forall j=1,\dots,n_m.
\label{M format}
\end{equation}
Similarly, $I_j^M$ denotes the set of indices of non-missing values in the $j$th column of $\boldsymbol{R}$,
$\boldsymbol{\widehat{U}}_{I_i^M}$ denotes the sub-matrix of $\boldsymbol{\widehat{U}}$ where rows $r \in I_j^M$ are selected, and $\boldsymbol{R}\big(I_j^M, j\big)$ is the column vector where rows $i \in I_i^M$ of the $j$th column of $\boldsymbol{R}$ are selected.

Step 4 (Iteration). Repeat Steps 2 and 3 until the stopping criterion is satisfied.

The sequence of achieved errors \eqref{object function} is proved monotone non-increasing and bounded below, hence this sequence converges \citep{lee2023randomly}.

\subsection{Core-Elements For Penalized Regression Splines}
\label{sec: core}
In large-scale data analysis tasks, various subsampling techniques such as uniform subsampling \citep{zhang2023model}, leverage-based subsampling \citep{ma2015leveraging,zhong2023model,shimizu2023improved}, and optimal information-based subsampling \citep{wang2019information,wu2024optimal,yu2023information} have been extensively studied and widely employed. These methods effectively reduce computational costs and have shown strong performance in practice.

Beyond these row-wise sampling approaches, \citetalias{li2024a} proposed the core-elements method for approximating the ordinary least squares (OLS) estimation in linear models.
To approximate the OLS estimation $\widehat{\boldsymbol{\beta}}=(\boldsymbol{X}^{\top} \boldsymbol{X})^{-1} \boldsymbol{X}^{\top} \y$, where $\boldsymbol{y} \in \mathbb{R}^n$ is the response vector and $\boldsymbol{X} \in \mathbb{R}^{n \times p}$ is the covariate matrix, the core-elements method constructs a sparse sketch $\boldsymbol{X}^* \in \mathbb{R}^{n \times p}$ of $\boldsymbol{X}$. Specifically, given a sampling budget \( s \in \mathbb{Z}_+ \) , let \( \boldsymbol{P} \in \mathbb{R}^{n \times p} \) be a binary matrix containing \( s \) ones and \( (np - s) \) zeros. 
The sparse sketch \( \boldsymbol{X}^* \) is then formed by \( \boldsymbol{X}^* = \boldsymbol{P} \odot \boldsymbol{X} \), where \( \odot \) denotes the element-wise product. Based on the sparse sketch, the core-element estimator $\widetilde{\boldsymbol{\beta}}=(\boldsymbol{X}^{\ast\top} \boldsymbol{X})^{-1} \boldsymbol{X}^{\ast\top} \y$ was proposed, motivated by the unbiasedness of the estimation.
This approach was later extended to nonparametric additive models by \citetalias{li2024b}, which approximated the penalized least squares (PLS) estimation of regression splines, taking the form 
\begin{equation}
\widetilde{\boldsymbol{\beta}}=\left(\boldsymbol{X}^{* \top} \boldsymbol{X}  +\boldsymbol{S}_{\lambda} \right)^{-1} \boldsymbol{X}^{* \top} \boldsymbol{y},
    \label{eq:CORE-PLS}
\end{equation}
where $\boldsymbol{S}_{\lambda}$ is a penalty term related to the smoothing parameter $\lambda$.

Subsequently, an upper bound on the estimator's variance is derived and approximately minimized, yielding the principle of core-element selection. Concretely, the sparse sketch $\boldsymbol{X}^*$ retains at most $s$ non-zero entries by keeping the $\lfloor s / p\rfloor$ largest absolute values in each column of $\boldsymbol{X}$ and zeroing out the rest. The resulting core-elements estimator provides theoretical approximation guarantees relative to the full-sample PLS and outperforms established subsampling methods in empirical studies.

\section{Methods}

\label{ch:3}
In Section~\ref{sec: core}, we introduced the core-elements approach designed for penalized regression splines. Notably, the iterative form of the alternating least squares (ALS) algorithm closely resembles that of penalized regression splines, suggesting the potential application of the core-elements method to ALS. Nonetheless, this extension is not straightforward mainly for three reasons.

Firstly, the formulation of penalized regression splines differs from that of ALS, necessitating the development of a new estimator tailored to the ALS framework. Additionally, ALS involves multiple regression computations, and mitigating computational complexity requires continuous sampling operations, which invariably incur significant time costs. This highlights the need for more efficient algorithms to optimize the sampling process.
Moreover, the core-elements method is inherently designed for sparse matrices, whereas the matrices involved in ALS are typically dense. This disparity underscores the need to evaluate the applicability and effectiveness of the core-elements approach within the context of dense matrices. Thus, directly applying the core-element algorithm to each regression step may inadvertently reduce the overall efficiency of the ALS algorithm, thereby necessitating careful algorithmic design.

In this section, we present our main algorithm. We first develop the core-elements estimation for ALS and introduce the principle of selecting core-elements motivated by approximately minimizing the upper bound for the mean squared Frobenius norm error of $\widetilde{\boldsymbol{U}}^{(t)}$ and $\widetilde{\boldsymbol{M}}^{(t)} $ in each iteration.

\textbf{Core-elements estimation.} Inspired by the formulation \eqref{eq:CORE-PLS},
 we propose an approximation for the ALS estimation \eqref{U format} and \eqref{M format} based on  sparse sketches $\boldsymbol{\widetilde{M}}_{I_i^U}^{(t)^{*}}$ and $\boldsymbol{\widetilde{U}}_{I_j^M}^{(t)^{*}}$ in each iteration, taking the form:
\begin{align}
    \widetilde{\boldsymbol{u}}_i^{(t+1)} & = (\boldsymbol{\widetilde{M}}_{I_i^U}^{(t)^{*^{\top}}} \boldsymbol{\widetilde{M}}_{I_i^U}^{(t)}+\lambda n_{u_i} \boldsymbol{E})^{-1} \boldsymbol{\widetilde{M}_{I_i^U}^{(t)^{*^{\top}}} }\boldsymbol{R}^{\top}\left(i, I_i^U\right), \quad \text{for}~~i =1,\dots,n_u, \label{eq:core_ui} \\
    \widetilde{\boldsymbol{m}}_j^{(t+1)} & = (\boldsymbol{\widetilde{U}}_{I_j^M}^{(t+1)^{*^{\top}}} \boldsymbol{\widetilde{U}}_{I_j^M}^{(t+1)}+\lambda n_{m_j} \boldsymbol{E})^{-1} \boldsymbol{\widetilde{U}}_{I_j^M}^{(t+1)^{*\top}} \boldsymbol{R}\big(I_j^M, j\big), \quad \text{for}~~j = 1,\dots,n_m, \label{eq:core_mj}
\end{align}
where $(t)$ denotes the $t$th iteration and $\boldsymbol{\widetilde{M}}_{I_i^U}^{(0)} = \boldsymbol{M}_{I_i^U}$.
We assume that $\boldsymbol{\widetilde{M}}_{I_i^U}^{(t)^{*^{\top}}} \boldsymbol{\widetilde{M}}_{I_i^U}^{(t)} + \lambda n_{u_i} \boldsymbol{E}$ and $\boldsymbol{\widetilde{U}}_{I_j^M}^{(t)^{*^{\top}}} \boldsymbol{\widetilde{U}}_{I_j^M}^{(t)} + \lambda n_{m_j} \boldsymbol{E}$ are of full rank. Based on the formulations in \eqref{eq:core_ui} and \eqref{eq:core_mj}, our goal is to iteratively find the sketches $\boldsymbol{\widetilde{M}}_{I_i^U}^{(t)^{*}}$ and $\boldsymbol{\widetilde{U}}_{I_j^M}^{(t)^{*}}$ that approximately minimize the expectation of the mean squared frobenius norm error (MSFE) of $\widetilde{\boldsymbol{U}}^{(t)} = (\widetilde{\boldsymbol{u}}_1^{(t)}, \dots, \widetilde{\boldsymbol{u}}_{n_u}^{(t)})^{\top}$ and $\widetilde{\boldsymbol{M}}^{(t)} = (\widetilde{\boldsymbol{m}}_1^{(t)}, \dots, \widetilde{\boldsymbol{m}}_{n_m}^{(t)})^{\top}$, respectively, as defined by:
\begin{align}
\operatorname{MFSE}\big(\widetilde{\boldsymbol{U}}^{(t)}\big) 
 &=  \mathbb{E}\big(\big\|\widetilde{\boldsymbol{U}}^{(t)}-\boldsymbol{U}^{(t)}\big\|_F^2\big) 
= \sum\limits_{i=1}^{n_u}\mathbb{E}\big(\|\widetilde{\boldsymbol{u}}_i^{(t)}-\boldsymbol{u}_i^{(t)}\|^2\big), \label{eq:mfse_u} \\
\operatorname{MFSE}\big(\widetilde{\boldsymbol{M}}^{(t)}\big) 
&=  \mathbb{E}\big(\big\|\widetilde{\boldsymbol{M}}^{(t)}-\boldsymbol{M}^{(t)}\big\|_F^2\big) 
= \sum\limits_{j=1}^{n_m}\mathbb{E}\big(\|\widetilde{\boldsymbol{m}}_j^{(t)}-\boldsymbol{m}_j^{(t)}\|^2\big),
\label{eq:mfse_m}
\end{align}
where $\boldsymbol{U}^{(t)} = (\boldsymbol{u}_1^{(t)}, \dots, \boldsymbol{u}_{n_u}^{(t)})^{\top}$ and $\boldsymbol{M}^{(t)} = (\boldsymbol{m}_1^{(t)}, \dots, \boldsymbol{m}_{n_m}^{(t)})^{\top}$ represent the ground truth values of $\boldsymbol{U}$ and $\boldsymbol{M}$ at the $t$th iteration, respectively.

Considering that it is challenging to directly minimize \eqref{eq:mfse_u} and \eqref{eq:mfse_m}, we provide upper bounds for $\operatorname{MFSE}\big(\widetilde{\boldsymbol{U}}^{(t)}\big)$ and   $\operatorname{MFSE}\big(\widetilde{\boldsymbol{M}}^{(t)}\big)$ in Proposition \ref{tm:1} and aim to minimize these upper bounds instead.

\begin{proposition}
    Let $(t)$ denote the $t$th iteration and $\boldsymbol{L}_{U_i}^{(t)} = \boldsymbol{\widetilde{M}}_{I_i^U}^{(t)} - \boldsymbol{\widetilde{M}}_{I_i^U}^{(t)^{*}}$. 
    Taylor expansions of $\mathbb{E}\big(\|\widetilde{\boldsymbol{u}}_i^{(t+1)}-\boldsymbol{u}_i^{(t+1)}\|^2\big)$ at $\boldsymbol{L}_{U_i}^{(t)}$ near the origin provide an upper bound for $\operatorname{MFSE}\big(\widetilde{\boldsymbol{U}}^{(t)}\big)$.

    \[\operatorname{MFSE}\big(\widetilde{\boldsymbol{U}}^{(t)}\big) \leq \boldsymbol{V}_u^{(t)} + \boldsymbol{B}_u^{(t)},
    \]
    where
    \[
    \boldsymbol{V}_u^{(t)} = \sum_{i=1}^{n_u} \sigma^2 \left\{ \left[1 + \mathcal{O}\left(\gamma_{u}^{(t)}\right)\right] \left( \|\boldsymbol{\widetilde{M}}_{I_i^U}^{(t)} \boldsymbol{D}_{U_i}^{(t)}\|_F^2 + \|\boldsymbol{D}_{U_i}^{(t)}\|_2^2 \|\boldsymbol{L}_{U_i}^{(t)}\|_F^2 \right) + \mathcal{O}\left(\gamma_{u}^{(t)}\right) \operatorname{Tr}(\boldsymbol{D}_{U_i}^{(t)}) \right\},
    \]
    and
    \[
    \boldsymbol{B}_u^{(t)} = \sum_{i=1}^{n_u} \left[1 + \mathcal{O}\left(\gamma_{u}^{(t)}\right)\right] \| \lambda n_{u_i} \boldsymbol{D}_{U_i}^{(t)} \boldsymbol{u}_i^{(t+1)} \|^2.
    \]
    Here, the terms $\boldsymbol{V}_u^{(t)}$ and $\boldsymbol{B}_u^{(t)}$ represent the upper bounds of the variance and the squared bias, respectively. The matrix $\boldsymbol{D}_{U_i}^{(t)}$ is defined as
    \[
    \boldsymbol{D}_{U_i}^{(t)} = \left( \boldsymbol{\widetilde{M}}_{I_i^U}^{(t)^{\top}} \boldsymbol{\widetilde{M}}_{I_i^U}^{(t)} + \lambda n_{u_i} \boldsymbol{E} \right)^{-1},
    \]
    and the spectral radius is given by
    \[
    \gamma_u^{(t)} = \| \boldsymbol{D}_{U_i}^{(t)} \boldsymbol{L}_{U_i}^{(t)^{\top}} \boldsymbol{\widetilde{M}}_{I_i^U}^{(t)} \|_2,
    \]
    which is assumed to satisfy $\gamma_u^{(t)} < 1$ to ensure the convergence of the matrix series.
    \label{proposition:mse_bound}
\end{proposition}

Similar to Proposition~\ref{proposition:mse_bound},  we also yield an upper bound for the $\operatorname{MFSE}\big(\widetilde{\boldsymbol{M}}^{(t)}\big)$ and the specific form of this upper bound can be found in the Appendix. Given the analogous forms of $\operatorname{MFSE}\big(\widetilde{\boldsymbol{U}}^{(t)}\big)$ and $\operatorname{MFSE}\big(\widetilde{\boldsymbol{M}}^{(t)}\big)$, we focus our analysis solely on $\operatorname{MFSE}\big(\widetilde{\boldsymbol{U}}^{(t)}\big)$. According to Proposition~\ref{proposition:mse_bound}, the upper bound of the $\operatorname{MFSE}\big(\widetilde{\boldsymbol{U}}^{(t)}\big)$ decreases as both $\|\boldsymbol{L}_{U_i}^{(t)}\|_F$ and the spectral radius $\gamma_u^{(t)}$ diminish. Specifically, the spectral radius $\gamma_u^{(t)}$ can be further bounded as follows: for $i = 1, \dots, n_u$,
\[
\gamma_u^{(t)} \leq \|\boldsymbol{D}_{U_i}^{(t)}\|_2 \|\boldsymbol{\widetilde{M}}_{I_i^U}^{(t)}\|_2 \|\boldsymbol{L}_{U_i}^{(t)}\|_2 \leq \|\boldsymbol{D}_{U_i}^{(t)}\|_2 \|\boldsymbol{\widetilde{M}}_{I_i^U}^{(t)}\|_2 \Big( n_f \max_{j \in \{1, \ldots, n_f\}} \boldsymbol{L}_{U_i}^{(t)^{(j) ^{\top}} }\boldsymbol{L}_{U_i}^{(t)^{(j)}} \Big)^{1/2},
\]
where $\boldsymbol{L}_{U_i}^{(t)^{(j)}}$ denotes the $j$th column of $\boldsymbol{L}_{U_i}^{(t)}$. This inequality indicates that a smaller maximum column norm of $\boldsymbol{L}_{U_i}^{(t)}$ leads to a smaller $\gamma_u^{(t)}$. Consequently, to minimize the upper bound of $\operatorname{MFSE}\big(\widetilde{\boldsymbol{U}}^{(t)}\big)$, it is essential to maintain both $\|\boldsymbol{L}_{U_i}^{(t)}\|_F$ and the column norms of $\boldsymbol{L}_{U_i}^{(t)}$ at minimal levels.

This requirement motivates a core-elements selection criterion similar to those proposed in \citetalias{li2024a} and \citetalias{li2024b}. Specifically, for each $i$, given a subsampling rate $r$, the sketch $\boldsymbol{\widetilde{M}}_{I_i^U}^{(t)^*}$ is constructed by retaining $\lfloor |I_i^U| \times r \rfloor$ elements with the largest absolute values in each column of $\boldsymbol{\widetilde{M}}_{I_i^U}^{(t)^*}$ and setting the remaining elements to zero. Intuitively, this approach ensures that $\boldsymbol{L}_{U_i}^{(t)}$ has approximately minimal column norms for each column. As a result, both $\|\boldsymbol{L}_{U_i}^{(t)}\|_F$ and $\|\boldsymbol{L}_{U_i}^{(t)}\|_2$ are approximately minimized, thereby achieving a relatively small upper bound for the $\operatorname{MFSE}\big(\widetilde{\boldsymbol{U}}^{(t)}\big)$ as stated in Proposition~\ref{proposition:mse_bound}.

Similarly, to minimize the upper bound of $\operatorname{MFSE}\big(\widetilde{\boldsymbol{M}}^{(t)}\big)$, we adopt an analogous core-elements selection criterion: for each $j$, given a subsampling rate $r$, the sketch $\boldsymbol{\widetilde{U}}_{I_j^M}^{(t)^*}$ is constructed by retaining $\lfloor |I_j^M| \times r \rfloor$ elements with the largest absolute values in each column of $\boldsymbol{\widetilde{U}}_{I_j^M}^{(t)^*}$ and setting the remaining elements to zero. This strategy also ensures that $\boldsymbol{L}_{M_j}^{(t)}$ maintains approximately minimal column norms for each column, thereby minimizing both $\|\boldsymbol{L}_{M_j}^{(t)}\|_F$ and $\|\boldsymbol{L}_{M_j}^{(t)}\|_2$. Consequently, the upper bound of the $\operatorname{MFSE}\big(\widetilde{\boldsymbol{M}}^{(t)}\big)$ is kept relatively small. 

We have also implemented a version using a fixed subsample size to ensure applicability across diverse experimental settings, and further compared row-wise and block-wise variants with the original column-wise scheme to demonstrate the advantages of column-wise core-elements sampling; see our Appendix for details.

According to the core-elements selection criterion abovementioned, Algorithm \ref{Alg:CES} provides the specific form of core-elements subsampling algorithm, which will be used in the acceleration of ALS.

 \vspace{0.8cm}
 \begin{algorithm}[H]
    \caption{Core-elements subsampling algorithm (CES)\label{Alg:CES}}
    \begin{algorithmic}[1]
        \State \textbf{Input:}  $\boldsymbol{X}=\left(x_{i j}\right) \in \mathbb{R}^{n \times p}$ , subsampling rate $r \in (0,1)$
        \State \textbf{Initialize}  $\boldsymbol{S}=(0) \in \mathbb{R}^{n \times p}$
        \State \textbf{For}~~$j=1, \ldots, p$  \textbf{do}
        \State \textbf{Let} $\mathcal{J}=\left\{i_1, \ldots, i_s\right\}$ be an index set , $s = r \times n$,  s.t. $\left\{\left|x_{i_q j}\right|\right\}_{q=1}^s$ are the $s$ largest ones among $\left\{\left|x_{i j}\right|\right\}_{i=1}^n$
        \State \textbf{Let} $s_{i_q j}=1, q=1, \ldots, s$
        \State \textbf{End for }
        \State \textbf{Let} $\boldsymbol{X}^*=\boldsymbol{S} \odot \boldsymbol{X}$ where $\odot$ represents the element-wise product
        \State \textbf{Return}  $\boldsymbol{X}^*$
    \end{algorithmic}
\end{algorithm}
\vspace{0.8cm}

\begin{figure}[h]
    \includegraphics[height=0.3\textheight,width=0.9\textwidth]{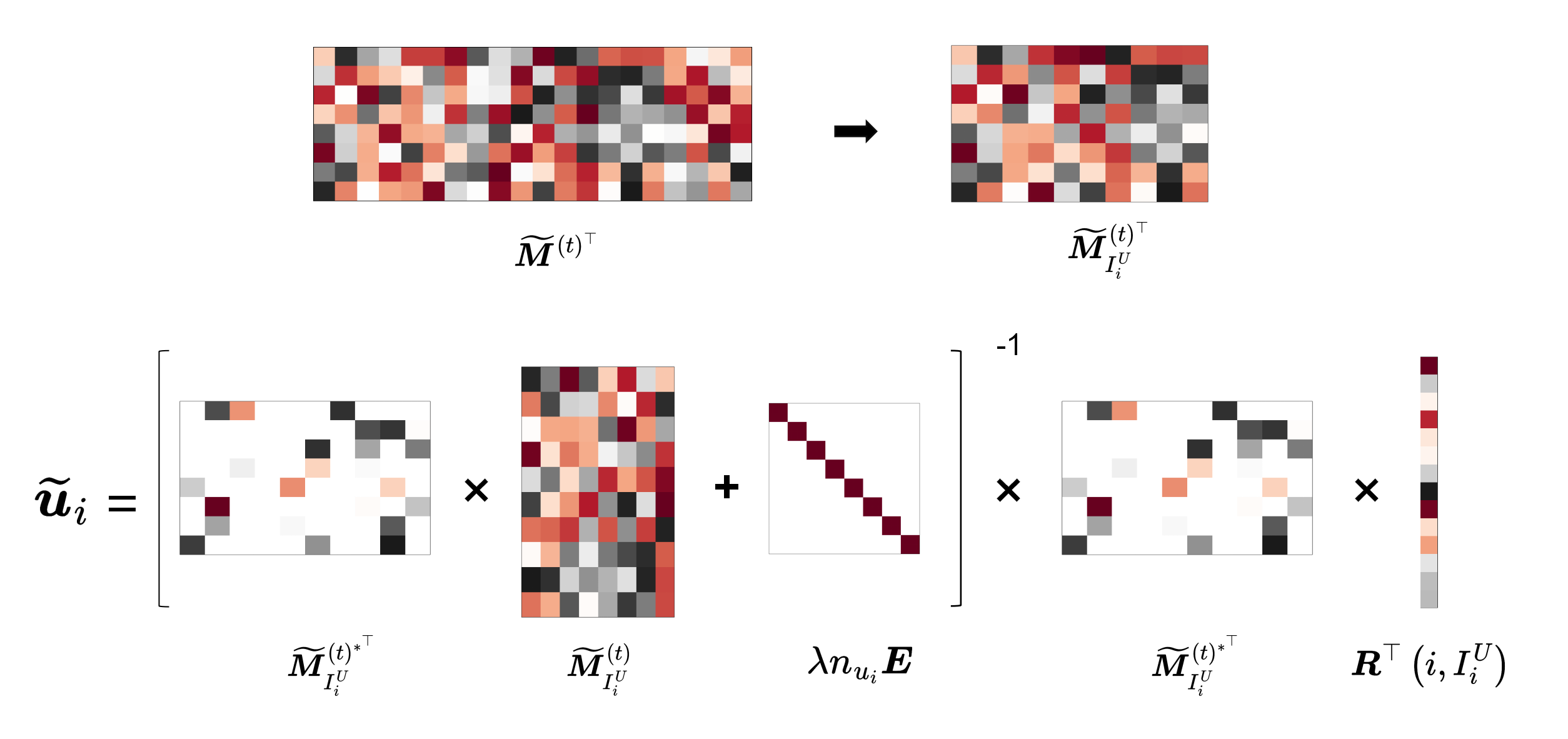}
    \centering
    \caption{Schematic of the flow of the  Core-ALS.}
    \label{Standard CORE ALS}
\end{figure}

Combining the abovementioned procedures, Algorithm \ref{Ag:Standard CORE ALS} summarizes the Core-ALS method, which uses the core-elements subsampling method to construct the sparse sketch for every regression. Schematic of the  Core-ALS method is shown in Fig.
\ref{Standard CORE ALS}.

\begin{algorithm}[H]
    \caption{Core-ALS \label{Alg:S-CORE-ALS}}
    \begin{algorithmic}[1]
        \State \textbf{Input:} rating matrix $\boldsymbol{R} \in \mathbb{R} ^{n_{u}\times n_{m}}$, implicit vector dimension $n_f$, subsampling rate $r$
        \State \textbf{Initialize} Matrices  $\boldsymbol{M}$ with ranks of $n_f$ and let $\boldsymbol{\widetilde{M}}^{(0)} = \boldsymbol{M}$
        \State \textbf{Repeat}
        \State \textbf{For}~~$ i = 1,\dots,n_u:$
        \State \quad Construct the sparse sketch:    $\boldsymbol{\widetilde{M}}_{I_i^U}^{(t)^{*}} = \operatorname{CES}(\boldsymbol{\widetilde{M}}_{I_i^U}^{(t)}, r)$
        \State  \quad  Update $ \widetilde{\boldsymbol{u}}_i^{(t+1)}$ with \eqref{eq:core_ui}
        \State \textbf{For}~~$ j = 1,\dots,n_m:$
        \State \quad Construct the sparse sketch:    $\boldsymbol{\widetilde{U}}_{I_j^M}^{(t+1)^{*}} = \operatorname{CES}(\boldsymbol{\widetilde{U}}_{I_j^M}^{(t+1)},r)$
        \State  \quad Update $ \widetilde{\boldsymbol{m}}_j^{(t+1)}$ with \eqref{eq:core_mj}
        \State \textbf{Until} \quad Convergence
        \State \textbf{Return}  \quad $\widetilde{\boldsymbol{U}}$ and $\widetilde{\boldsymbol{M}}$
    \end{algorithmic}
    \label{Ag:Standard CORE ALS}
\end{algorithm}

In Algorithm \ref{Ag:Standard CORE ALS}, selecting elements of the $\widetilde{\boldsymbol{U}}^{(t)}$ or $\widetilde{\boldsymbol{M}}^{(t)}$ slices before each regression step can lead to some time loss during execution. To mitigate this issue, we optimized the sampling algorithm by employing Partial Quicksort \citep{martinez2004partial} to sort the elements based on their magnitudes (see Appendix for details). This optimization ensures that our sampling process remains nearly lossless in the case of large-scale matrices. We reported sorting costs in the Appendix to show the efficiency of our sorting
procedure. Furthermore, our method provides a distinct advantage over alternative techniques that require computing sampling probabilities, such as leverage score-based sampling, which often incur substantial computational costs.

\begin{figure}[h]
\includegraphics[height=0.3\textheight,width=0.9\textwidth]{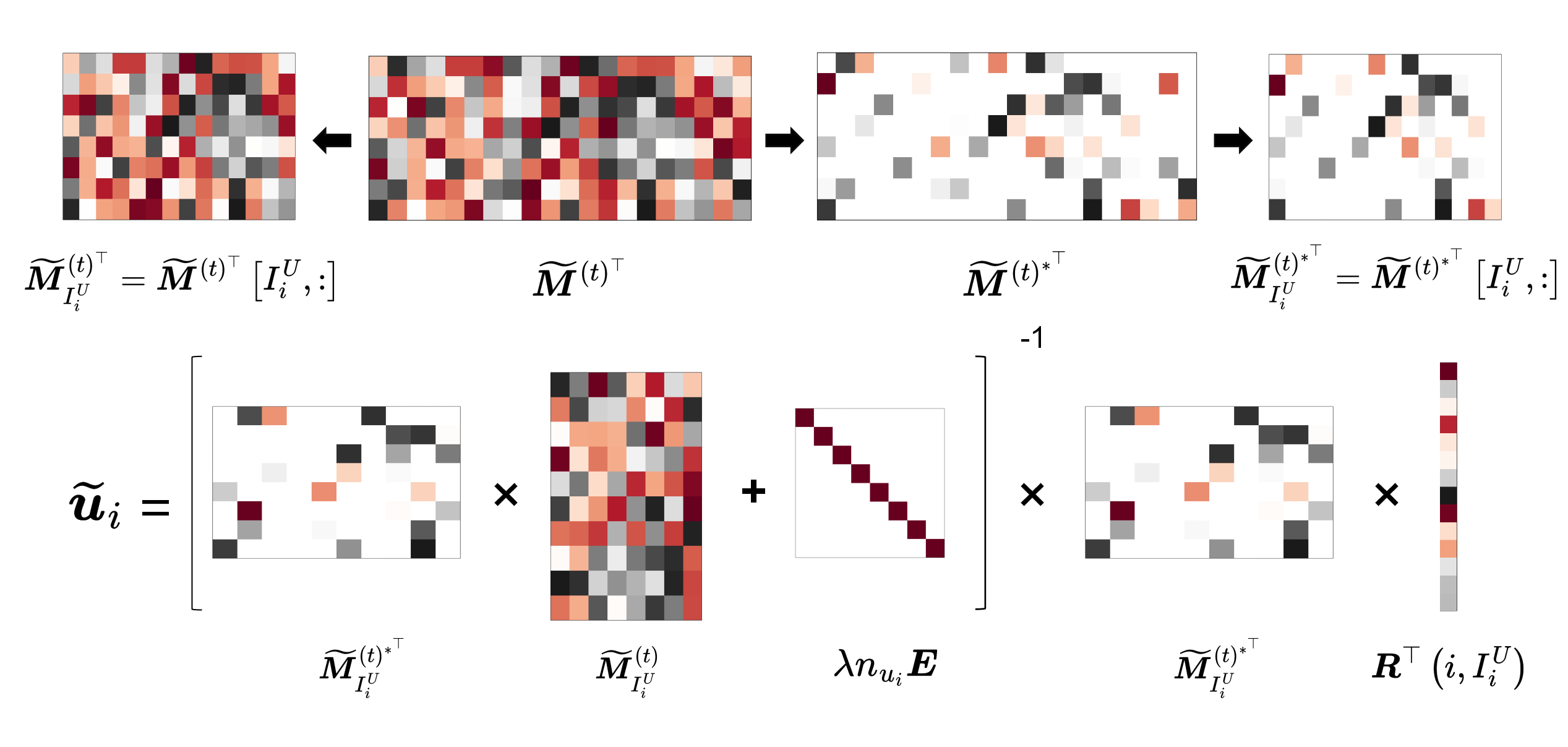}
    \centering
    \caption{Fast variant of the  Core-ALS.}
    \label{fast CORE ALS}
\end{figure}

In addition, we introduce a fast variant of the algorithm, which performs sampling on $\widetilde{\boldsymbol{U}}^{(t)}$ and $\widetilde{\boldsymbol{M}}^{(t)}$ before each iteration rather than repeatedly sampling on $\boldsymbol{\widetilde{U}}_{I_j^M}^{(t)}$ and $\boldsymbol{\widetilde{M}}_{I_i^U}^{(t)}$. While this approach does not guarantee exact adherence to the specified sampling ratio, it ensures that the time complexity does not exceed that of the original algorithm. Our experimental results suggest that this variant significantly lowers the sampling overhead while maintaining relatively high accuracy. Schematic of the fast variant of the Core-ALS is shown in Fig.
\ref{fast CORE ALS} and a detailed description of this method is provided in the Appendix.

Moreover, the core-elements estimation proposed here is specifically designed for the explicit-feedback ALS. To extend core-elements estimation to the implicit-feedback setting, we combine it with the implicit-feedback variant of ALS in the Appendix and provide the corresponding experiments.


\section{Theoretical Results}\label{sec:theory}
In this section, we first demonstrate that the core-elements estimation achieves the $(1 + \epsilon)$-approximation
w.r.t. the full sample estimation \eqref{U format} and \eqref{M format}. Then, we show the convergence of the algorithm. Finally, the time complexity of the two proposed algorithms is given.
Technical proofs are provided in the Appendix.
\label{ch:4}
\subsection{Approximation Guarantee}
Theorems \ref{tm:1}  and  \ref{tm:2} provides  non-asymptotic relative error bounds for the proposed core-elements estimation $\widetilde{\boldsymbol{u}}_i$ and $\widetilde{\boldsymbol{m}}_j$ in each iteration, respectively.

\begin{theorem}\label{tm:1}
Let $\boldsymbol{\widetilde{M}}_{I_i^U}^{(t)^{*}}$ be the sparse sketch of $\boldsymbol{\widetilde{M}}_{I_i^U}^{(t)}$, and recall $\widetilde{\boldsymbol{u}}_i^{(t+1)}$ defined in \eqref{eq:core_ui}. Suppose 
\[
\big\|\boldsymbol{\widetilde{M}}_{I_i^U}^{(t)} - \boldsymbol{\widetilde{M}}_{I_i^U}^{(t)^{*}}\big\|_2 \leq \epsilon_m' \big\|\boldsymbol{\widetilde{M}}_{I_i^U}^{(t)}\big\|_2,
\]
where
\[
0 < \epsilon_m' \leq \frac{1}{c_m^{(t)}} \left[1 + \frac{c_m^{(t)} + 1}{(\sqrt{1 + \epsilon_m} - 1)\operatorname{RSSE}(\widehat{\boldsymbol{u}}_i^{(t+1)})}\right]^{-1}.
\]
Under these conditions, we have
\begin{equation}
   \big\|\boldsymbol{R}^{\top}(i, I_i^U) - \boldsymbol{\widetilde{M}}_{I_i^U}^{(t)}\widetilde{\boldsymbol{u}}_i^{(t+1)}\big\|^2 \leq (1+\epsilon_m)\big\|\boldsymbol{R}^{\top}(i, I_i^U) - \boldsymbol{\widetilde{M}}_{I_i^U}^{(t)}\widehat{\boldsymbol{u}}_i^{(t+1)}\big\|^2.
\label{eq:tm1} 
\end{equation}

In \eqref{eq:tm1}, $c_m^{(t)}=\big\|\boldsymbol{\widetilde{M}}_{I_i^U}^{(t)}\big\|_2^2\Big\|\left(\boldsymbol{\widetilde{M}}_{I_i^U}^{(t)^{\top}} \boldsymbol{\widetilde{M}}_{I_i^U}^{(t)}+\lambda n_{u_i} \boldsymbol{E}\right)^{-1}\Big\|_2$ 
and  $\operatorname{RSSE}\left(\widehat{\boldsymbol{u}}_i^{(t+1)}\right)=\big\|\boldsymbol{R}^{\top}\left(i, I_i^U\right)-\boldsymbol{\widetilde{M}}_{I_i^U}^{(t)} \widehat{\boldsymbol{u}}_i^{(t+1)}\big\| /\big\|\boldsymbol{R}^{\top}\left(i, I_i^U\right)\big\|$ 
is the relative sum of squares error (RSSE) of the full sample estimation $\widehat{\boldsymbol{u}}_i^{(t+1)}$.
\end{theorem}

\begin{theorem}\label{tm:2}
Let $\boldsymbol{\widetilde{U}}_{I_j^M}^*$ be the sparse sketch of $\boldsymbol{\widetilde{U}}_{I_j^M}$, and $\widetilde{\boldsymbol{m}}_j$ be defined in \eqref{eq:core_mj}. 
Under conditions similar to those in Theorem \ref{tm:1}, we also have
\begin{equation}
    \big\|\boldsymbol{R}^{\top}(j, I_j^M) - \boldsymbol{\widetilde{U}}_{I_j^M}^{(t+1)}\widetilde{\boldsymbol{m}}_j^{(t+1)}\|^2 \leq (1+\epsilon_u)\big\|\boldsymbol{R}^{\top}(j, I_j^M) - \boldsymbol{\widetilde{U}}_{I_j^M}^{(t+1)}\widehat{\boldsymbol{m}}_j^{(t+1)}\big\|^2.
    \label{eq:tm2}
\end{equation}

The complete version of this theorem can be found in the Appendix.
\end{theorem}

Theorems \ref{tm:1} and \ref{tm:2} indicate that to achieve the $(1+\epsilon)$-approximation, Algorithm \ref{Ag:Standard CORE ALS} requires sketches $\boldsymbol{\widetilde{M}}_{I_i^U}^{(t)^*}$ and $\boldsymbol{\widetilde{U}}_{I_j^M}^{(t+1)^*}$ such that the ratios ${\big\|\boldsymbol{\widetilde{M}}_{I_i^U}^{(t)}-\boldsymbol{\widetilde{M}}_{I_i^U}^{(t)^*}\big\|_2 }/{\big\|\boldsymbol{\widetilde{M}}_{I_i^U}^{(t)}\big\|_2}$ and $\big\|\boldsymbol{\widetilde{U}}_{I_j^M}^{(t+1)}-\boldsymbol{\widetilde{U}}_{I_j^M}^{(t+1)^*}\big\|_2 /\big\|\boldsymbol{\widetilde{U}}_{I_j^M}^{(t+1)}\big\|_2$ are $O\left(\epsilon^{1 / 2}\right)$, respectively. 

\subsection{Convergence Guarantee}
\begin{theorem}
We use the superscript $(t)$ to represent the $t$th step of the iteration. $\boldsymbol{\widetilde{M}}_{I_i^U}^{(t)^*}$ is the sparse sketch of $\boldsymbol{\widetilde{M}}_{I_i^U}^{(t)}$ and $\widetilde{\boldsymbol{u}}_i^{(t+1)}$ is defined by \eqref{eq:core_ui}. $\boldsymbol{\widetilde{U}}_{I_j^M}^{(t+1)^*}$ is the sparse sketch of $\boldsymbol{\widetilde{U}}_{I_j^M}^{(t+1)}$ and $\widetilde{\boldsymbol{m}}_j^{(t+1)}$ is defined by \eqref{eq:core_mj}.
 $\epsilon_m$, $\epsilon_u$, $\epsilon^{\prime}_{m}$, $\epsilon^{\prime}_{u}$ are defined in Theorems \ref{tm:1} and \ref{tm:2}, respectively.

Suppose the full sample estimator's convergence rate is $C$. When $C$ satisfies $C\leq \min\{1/(1+\epsilon_u),1/(1+\epsilon_m)\}$, $ \epsilon^{\prime}_{m}$ and $ \epsilon^{\prime}_{u}$ always satisfy conditions in Theorems \ref{tm:1} and \ref{tm:2}, i.e., are $O(\epsilon^{1/2})$.

We have
\begin{align}
    \label{eq:conv1}
    \sum_{j=1}^{n_m}\big\|\boldsymbol{R}^{\top}\left(j, I_j^M\right)-\boldsymbol{\widetilde{U}}_{I_j^M}^{(t+1)} \widetilde{\boldsymbol{m}}_j^{(t+1)}\big\|^2 &\leq \sum_{i=1}^{n_u}\big\|\boldsymbol{R}^{\top}\left(i, I_i^U\right)-\boldsymbol{\widetilde{M}}_{I_i^U}^{(t)} \widetilde{\boldsymbol{u}}_i^{(t+1)}\big\|^2,\\
    \label{eq:conv2}
    \sum_{i=1}^{n_u}\big\|\boldsymbol{R}^{\top}\left(i, I_i^U\right)-\boldsymbol{\widetilde{M}}_{I_i^U}^{(t)} \widetilde{\boldsymbol{u}}_i^{(t+1)}\big\|^2 &\leq \sum_{j=1}^{n_m}\big\|\boldsymbol{R}^{\top}\left(j, I_j^M\right)-\boldsymbol{\widetilde{U}}_{I_j^M}^{(t)} \widetilde{\boldsymbol{m}}_j^{(t)}\big\|^2.
\end{align}

Combining the equation \eqref{eq:conv1} and equation \eqref{eq:conv2}, we have
\begin{equation*}
\sum_{j=1}^{n_m}\big\|\boldsymbol{R}^{\top}\left(j, I_j^M\right)-\boldsymbol{\widetilde{U}}_{I_j^M}^{(t+1)} \widetilde{\boldsymbol{m}}_j^{(t+1)}\big\|^2 \leq
\sum_{j=1}^{n_m}\big\|\boldsymbol{R}^{\top}\left(j, I_j^M\right)-\boldsymbol{\widetilde{U}}_{I_j^M}^{(t)} \widetilde{\boldsymbol{m}}_j^{(t)}\big\|^2,
\end{equation*}
i.e.,
\begin{equation*}
    {\mathcal{L}_\lambda^{r e g}(\boldsymbol{R}, \boldsymbol{U}, \boldsymbol{M})}^{(t+1)}\leq {\mathcal{L}_\lambda^{r e g}(\boldsymbol{R}, \boldsymbol{U}, \boldsymbol{M})}^{(t)}.
\end{equation*}

Under these conditions, the error function is monotonically decreasing, so the algorithm converges.
    \label{tm:3}
\end{theorem}
Theorem \ref{tm:3} indicates that as long as the full sample estimator's convergence rate is small enough and the sampled matrix closely approximates the original matrix, Core-ALS achieves convergence.

\subsection{Time Complexity Analysis}
Theorem \ref{complex1}  provides specific time complexity for the proposed core-elements ALS method.

\begin{theorem}
Suppose $r$ is the subsampling rate, $\text{nnz}(R)$ is the number of non-missing values in the rating matrix $\boldsymbol{R}$.
For the Core-ALS method, each step of updating $\boldsymbol{U}$ takes 
  \begin{equation*}
    O\left(n_f\left(n n z(R) \times \textcolor{red}{\left(r n_f\right)}+ n_f^2 n_u\right)\right).  
\end{equation*}
while each step of updating $\boldsymbol{M}$ takes 
\begin{equation*}
    O\left(n_f\left(n n z(R) \times\textcolor{red}{\left(r n_f\right)} + n_f^2 n_m\right)\right).
\end{equation*}
If the Core-ALS method takes a total of $n_t$ rounds to stop,
it runs in time 
\begin{equation}
    O\left(n_f\left(n n z(R) \times\textcolor{red}{\left(r n_f\right)}+ n_f^2 n_m+n_f^2 n_u\right) n_t\right).
\end{equation}
\label{complex1}
\end{theorem}
Theorem \ref{complex1} shows that Core-ALS substantially reduces the computational cost of ALS. 
When the sampling probability is sufficiently small, the overall time complexity of the algorithm can reach $O\left(n_f^3\left(n_m+n_u\right) n_t\right)$.


\section{Simulation Studies}
\label{ch:5}
In this section, we evaluate the performance of the Core-ALS method using synthetic data. We use CORE to refer to the estimator in Algorithm \ref{Ag:Standard CORE ALS}. For comparison, we consider
several state-of-the-art subsampling methods including uniform subsampling (UNIF), basic
leverage subsampling (BLEV) \citep{ma2015leveraging,WOS:000458973703072}. We have also compared with two production-grade speed-up baselines, SparkALS \citep{winlaw2015algorithmic} and SGD-ALS, in the Appendix.
All experiments were implemented using the R programming language on a server with 256 GB RAM and 64 cores Intel ${ }^{\circledR}$ Xeon ${ }^{\circledR}$ Gold 5218 CPU.

\subsection{Estimation Accuracy under Different Parameter Settings}

To construct a low-rank rating matrix $\boldsymbol{R}$, we first generate two low-rank factor matrices $\boldsymbol{U}$ and $\boldsymbol{M}$ and then form an initial low-rank matrix $\boldsymbol{R}^{o} = \boldsymbol{U}\boldsymbol{M}^{\top}$ before sparsification. The factor matrices $\boldsymbol{U}$ and $\boldsymbol{M}$ are generated from one of the following widely used multivariate distributions:

$\mathbf{D1}$. multivariate normal distribution, $N(\boldsymbol{0}, \boldsymbol{\Sigma})$;

$\mathbf{D2}$. multivariate log-normal distribution, $L N(\boldsymbol{0}, \boldsymbol{\Sigma})$;

$\mathbf{D3}$. multivariate t-distribution with 4 degrees of freedom, $t_4(\boldsymbol{0}, \boldsymbol{\Sigma})$,
where $\boldsymbol{\Sigma}=\left(\sigma_{i j}\right) \in \mathbb{R}^{p \times p}$ is a covariance matrix with $\sigma_{i j}=0.6^{|i-j|}$ for $i, j=1, \ldots, p$.

After constructing the initial low-rank matrix $\boldsymbol{R}^{o}$, we add Gaussian noise $\varepsilon_{ij} \sim N(0,1)$ to each entry to introduce stochastic perturbations. Subsequently, we randomly remove a proportion of the entries to achieve a prescribed sparsity ratio $\alpha$. Specifically, we randomly select $(1-\alpha) \times 100\%$ of the entries and set them to \texttt{NaN}. We consider $\alpha \in \{0.4,0.5,0.6,0.7\}$, which correspond to rating matrices denoted by $\mathbf{R1}$ through $\mathbf{R4}$, respectively. Note that $\mathbf{R1}$ represents a relatively dense matrix, while $\mathbf{R4}$ is relatively sparse.

We fix the size of the rating matrix as $(n_u,n_m) = (3600,3600)$ and choose the implicit vector dimension $n_f = 60$. For the three subsampling methods, i.e., UNIF, BLEV, and CORE, we select a subsampling rate $r \in \{0.1, 0.15, 0.2, 0.25\}$ for each regression problem. Under these chosen subsampling rates, all three methods maintain theoretically comparable time complexity.

Due to the sparsity of the rating matrix, we take the position of the non-missing value as the empirical set and the position of the missing value as the prediction set. We calculate the empirical relative mean squared error (ReMSE) for each of the estimators based on one hundred replications, i.e.,
\begin{equation*}
    \mathrm{ReMSE}(\boldsymbol{\widetilde{R}})=\frac{1}{100} \sum_{n=1}^{100}  \frac{\sqrt{\sum_{(i, j) \in I}\left(r_{i j}-\widetilde{r}_{i j}\right)^2}}{\sqrt{\sum_{(i, j) \in I} r_{i j}{ }^2}}, 
\end{equation*}
where $I$ represents the set of non-missing value locations, $\widetilde{r}$ represents the elements of the approximate matrix $\boldsymbol{\widetilde{R}}$ obtained by the ALS estimator and calculate the prediction relative MSE (P-ReMSE) for each of the estimators based on one hundred replications, i.e.,
\begin{equation*}
\text{P-ReMSE}(\boldsymbol{\widetilde{R}})=\frac{1}{100} \sum_{n=1}^{100}  \frac{\sqrt{\sum_{(i, j) \in I^{c}}\left(r_{i j}-\widetilde{r}_{i j}\right)^2}}{\sqrt{\sum_{(i, j) \in I^c} r_{i j}{ }^2}}, 
\end{equation*}
where $I^c$ represents the set of missing value locations. 

\begin{figure}[h]
    \centering

    \begin{minipage}[b]{0.496\textwidth}
        \centering
        \includegraphics[width=\linewidth]{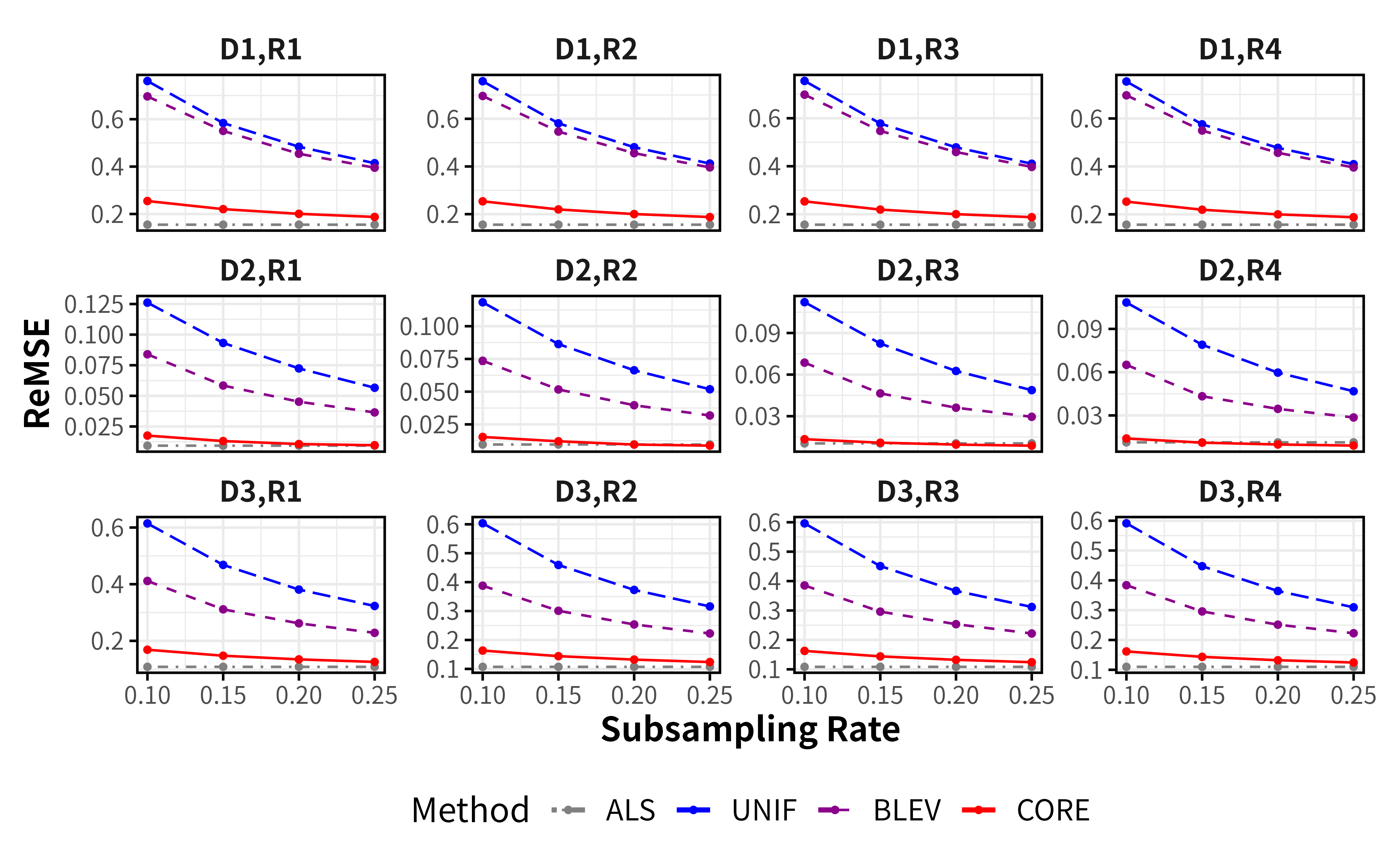}
        \caption*{\footnotesize(a) ReMSE for different D and R.}  

    \end{minipage}
    \begin{minipage}[b]{0.496\textwidth}
        \centering
        \includegraphics[width=\linewidth]{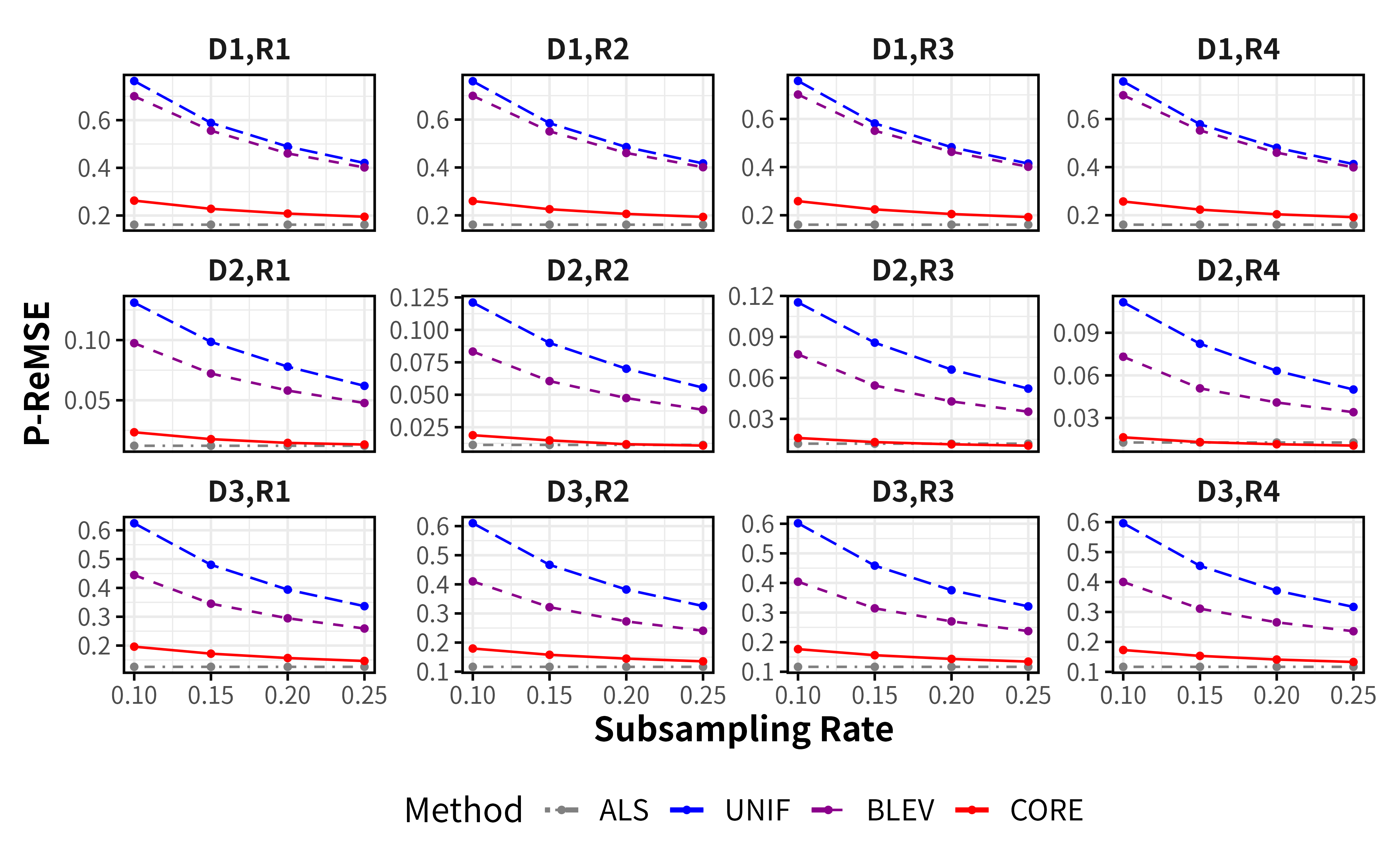}
        \caption*{\footnotesize(b) P-ReMSE for different D and R.}

    \end{minipage}


    \begin{minipage}[b]{0.496\textwidth}
        \centering
        \includegraphics[width=\linewidth]{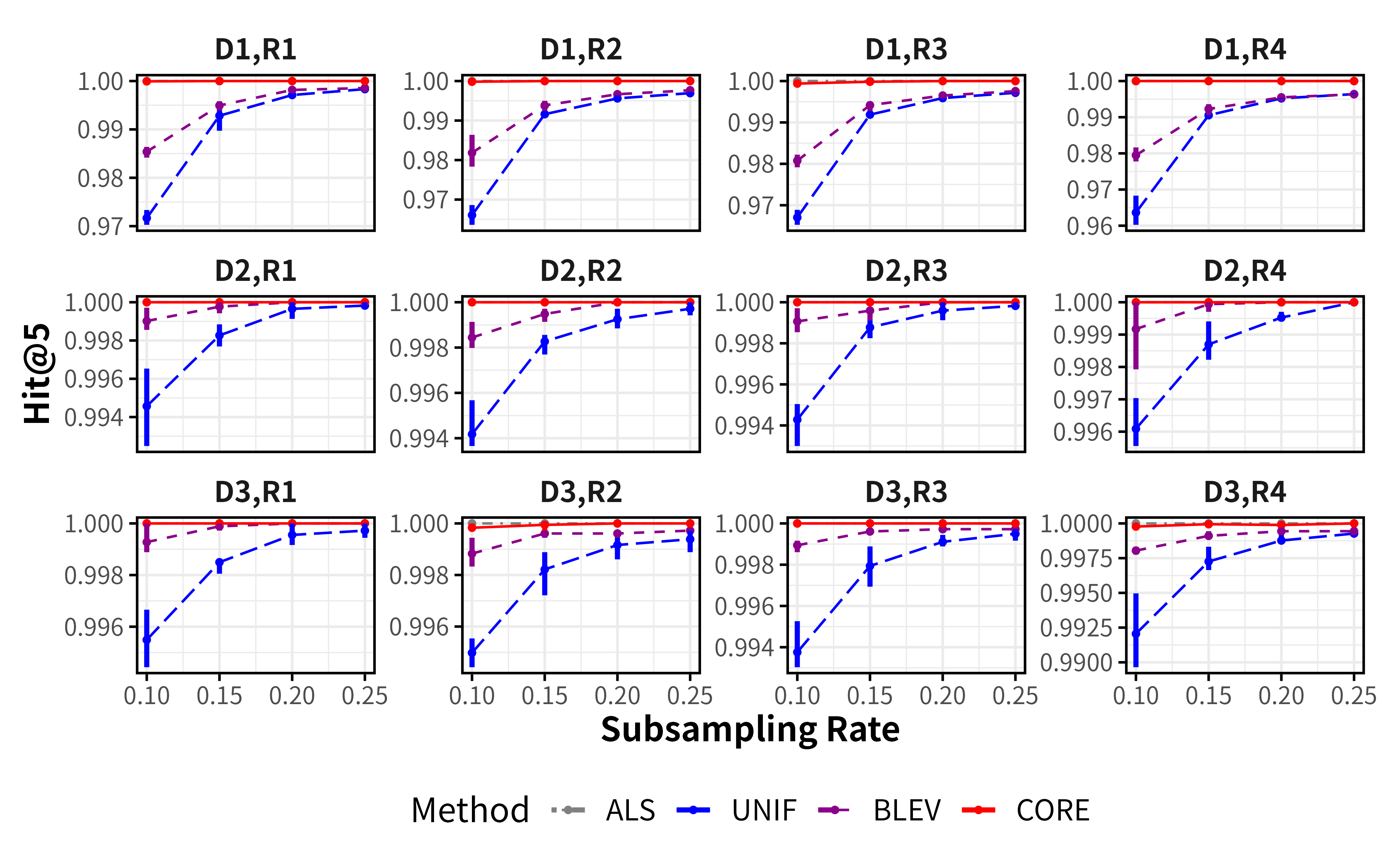}
        \caption*{\footnotesize(c) Hit@5 for different D and R.}

    \end{minipage}
    \begin{minipage}[b]{0.496\textwidth}
        \centering
        \includegraphics[width=\linewidth]{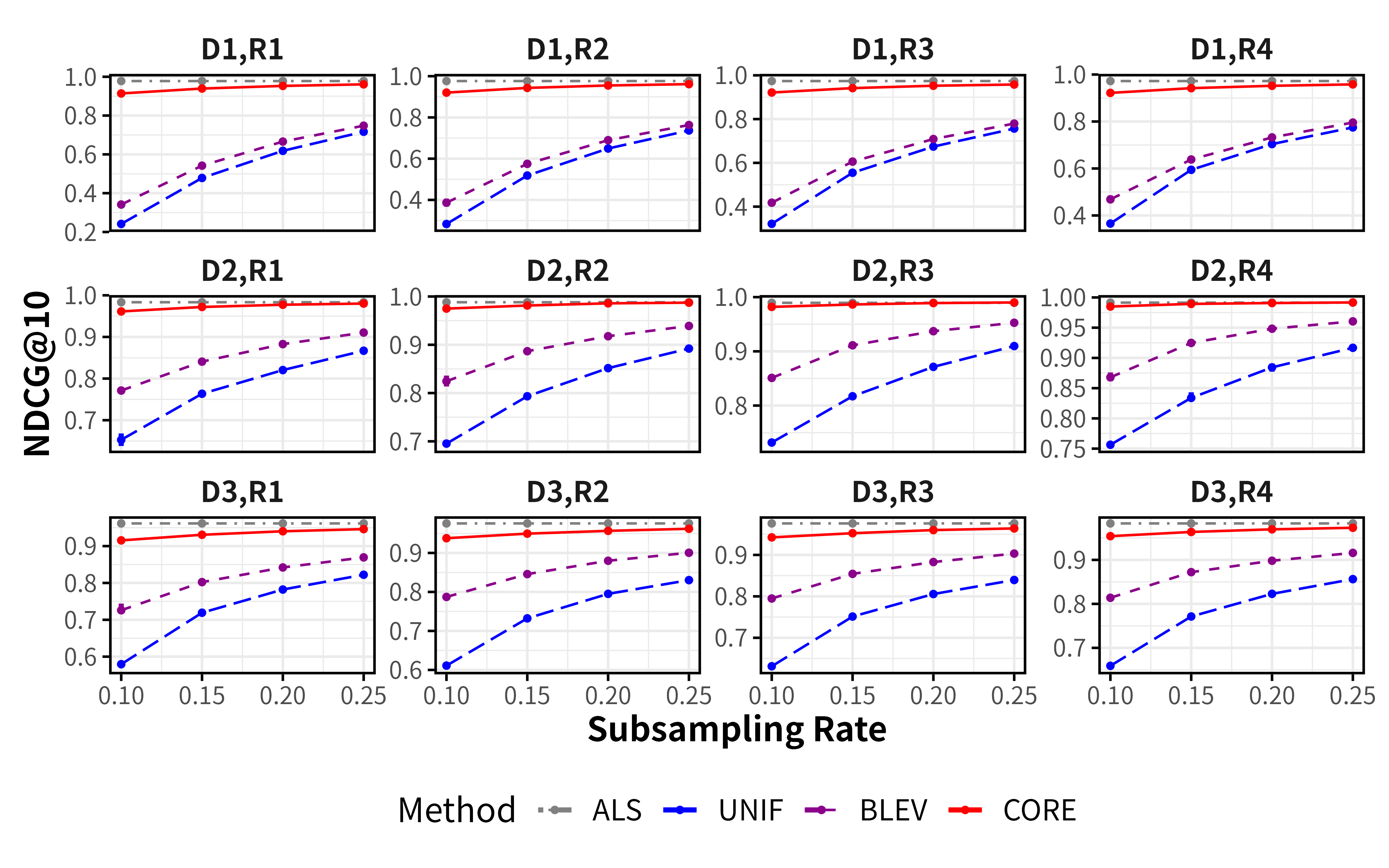}
        \caption*{\footnotesize(d) NDCG@10 for different D and R.}

    \end{minipage}


    \caption{Performance of four methods under different distributions and densities.}
    \label{fig:simu_DR}
\end{figure}

We also evaluate two widely-used metrics in recommender systems: Hit@5 and NDCG@10 (score-based version). Hit@5 measures whether the ground-truth item appears in the top-5 recommendations. NDCG@10 considers both the relevance and rank of recommended items. Our setting is explicit-feedback ALS for matrix reconstruction without a temporal dimension; recommendations are produced by ranking predicted ratings for unrated items. In contrast, classical leave-one-out or temporal split strategies were designed primarily for implicit-feedback scenarios or time-ordered recommendation scenarios. Randomly selecting a rating from the test set as the leave-one-out target would not reliably represent an item that should be recommended in this setting. Therefore, we consider the top $95\%$ of scores in the test set as the items that are actually recommended. To demonstrate robustness, we also included complementary experiments under the leave-one-out setting in the Appendix. The results of four metrics versus different subsample sizes are shown in Fig. \ref{fig:simu_DR}.

In Fig. \ref{fig:simu_DR}, we observe that both MSE and PMSE w.r.t. all estimators decrease as $r$ increases. We also observe that CORE consistently outperforms all other methods by a large margin on both Hit and NDCG metrics.
This observation demonstrates that the proposed estimator achieves superior accuracy compared to existing methods by effectively leveraging the information embedded in the rating matrix. In particular, the core-elements approach ensures that, at each iteration, the estimator remains approximately unbiased and maintains an approximately minimized estimation variance, thereby conferring a clear advantage over competing techniques.

\begin{figure}[h]
    \centering

    \begin{minipage}[b]{0.496\textwidth}
        \centering
        \includegraphics[width=\linewidth]{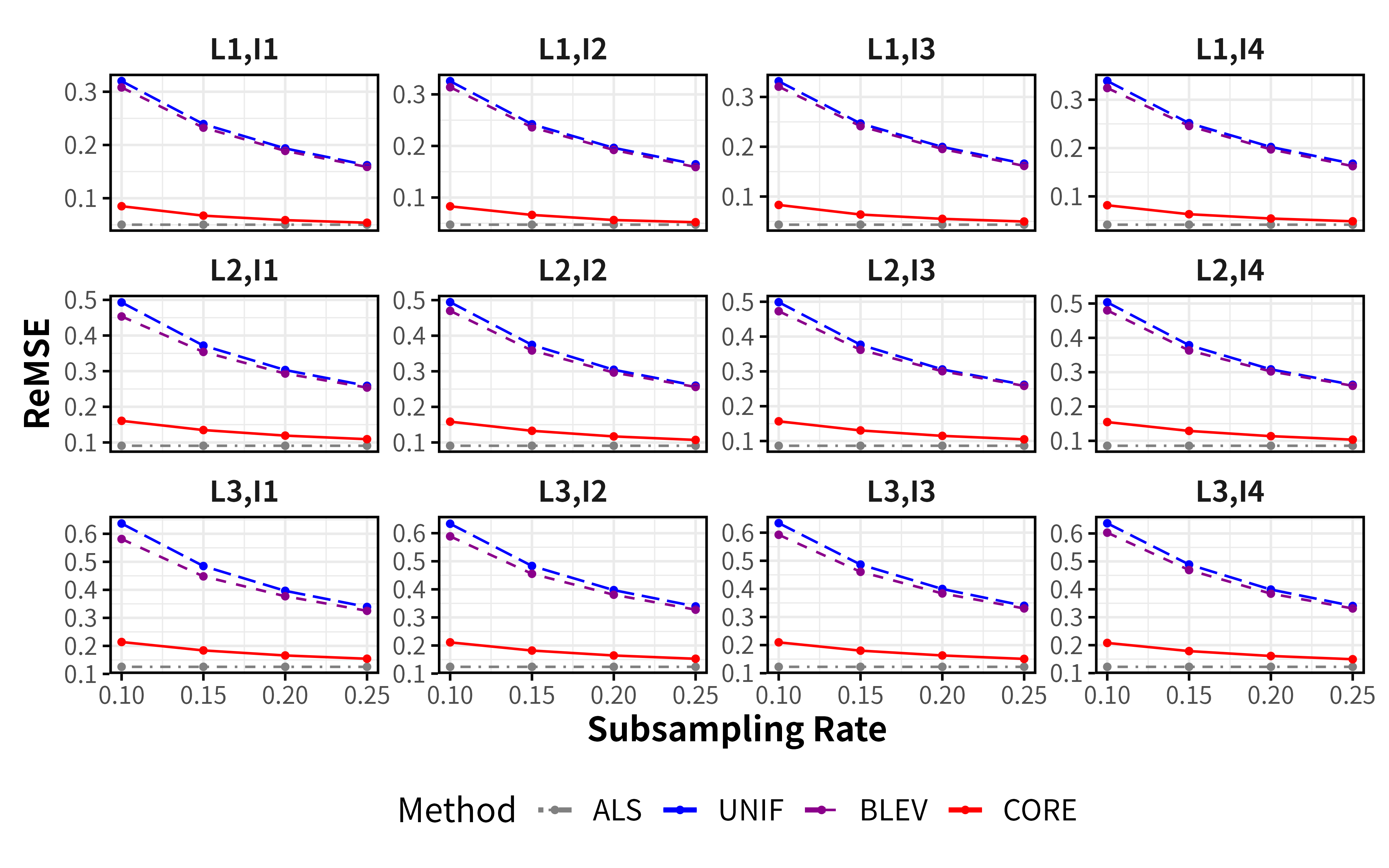}
        \caption*{\footnotesize(a) ReMSE for different L and I.}  
    \end{minipage}
    \begin{minipage}[b]{0.496\textwidth}
        \centering
        \includegraphics[width=\linewidth]{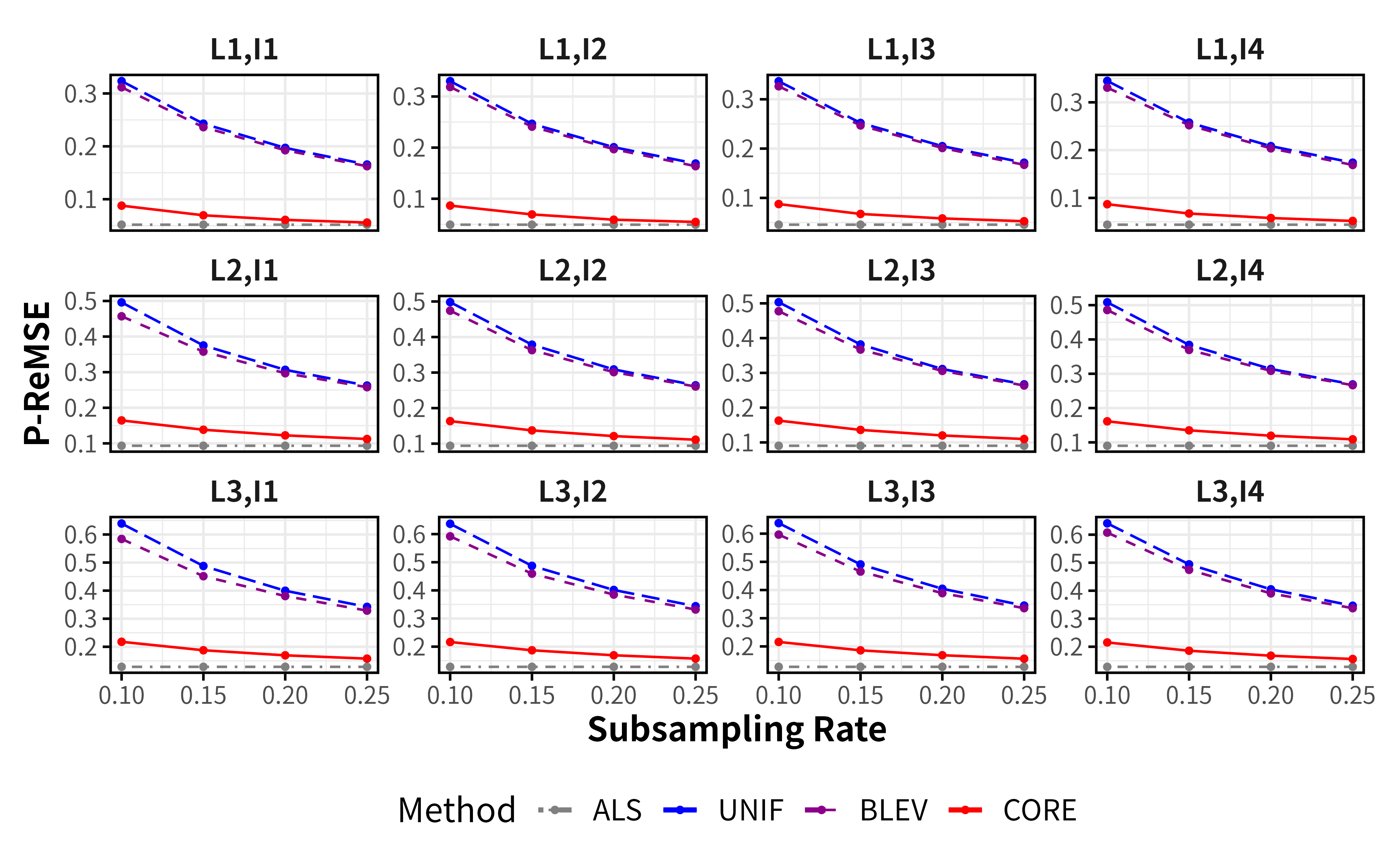}
        \caption*{\footnotesize(b) P-ReMSE for different L and I.}
    \end{minipage}


    \begin{minipage}[b]{0.496\textwidth}
        \centering
        \includegraphics[width=\linewidth]{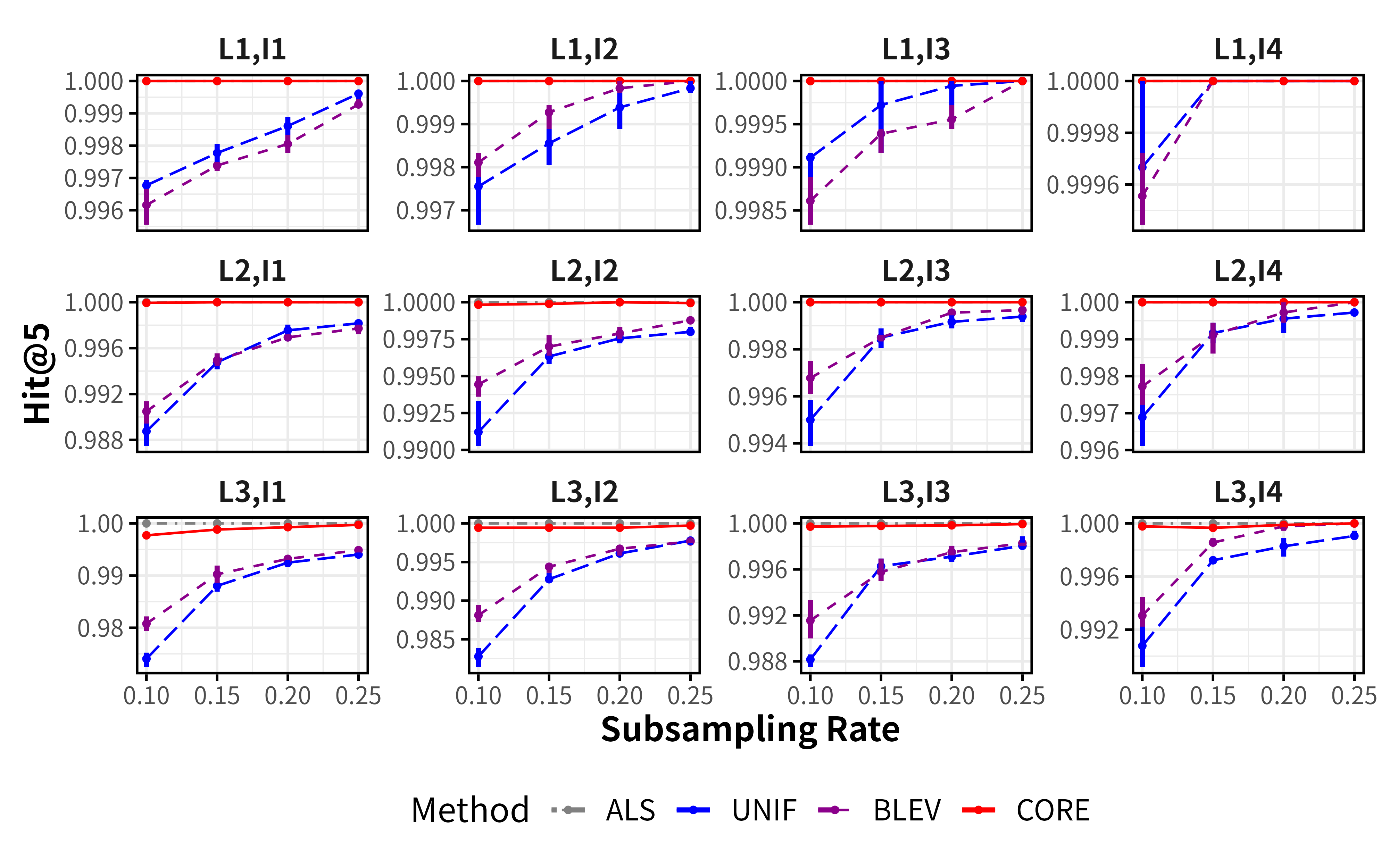}
        \caption*{\footnotesize(c) Hit@5 for different L and I.}
    \end{minipage}
    \begin{minipage}[b]{0.496\textwidth}
        \centering
        \includegraphics[width=\linewidth]{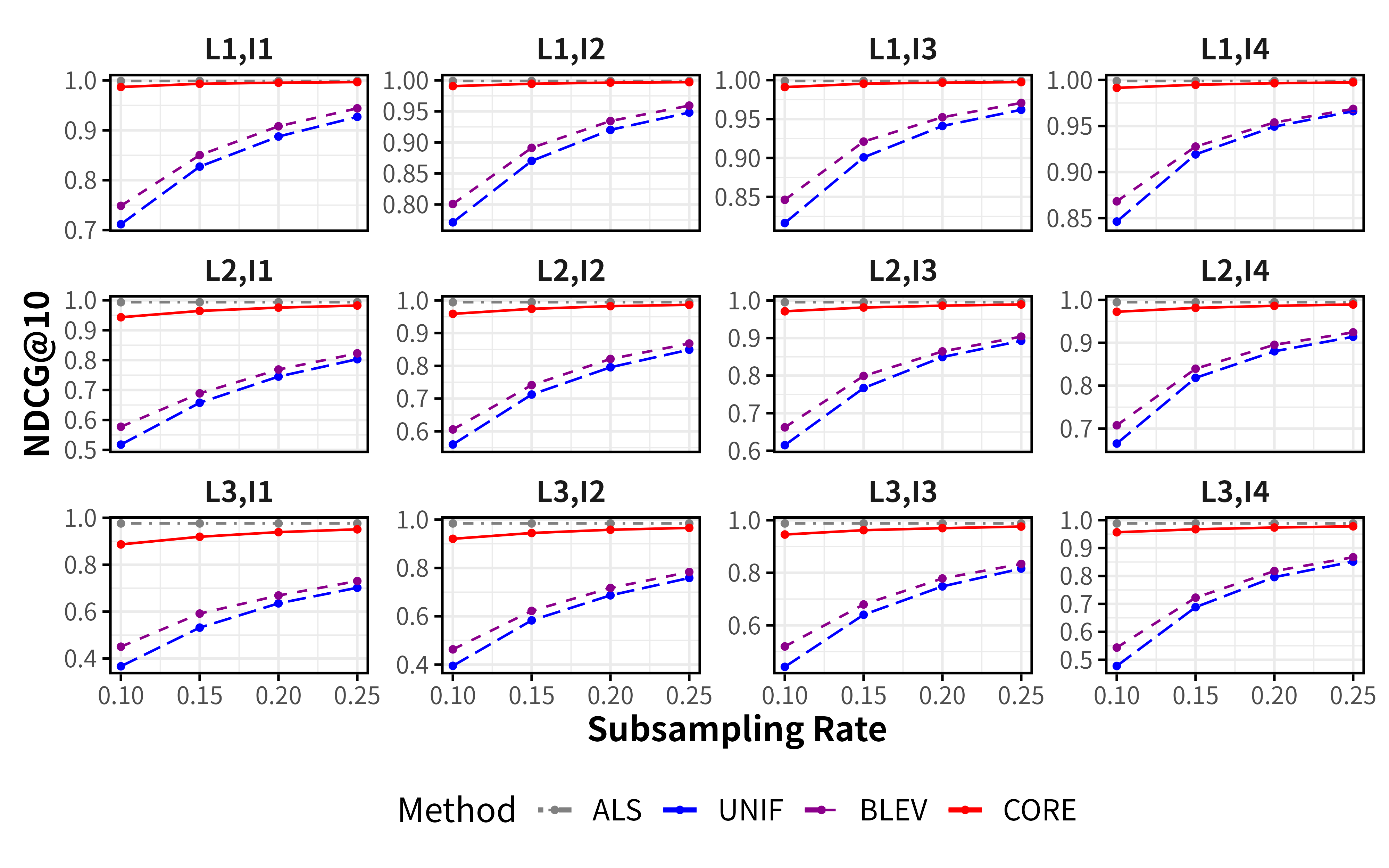}
        \caption*{\footnotesize(d) NDCG@10 for different L and I.}
    \end{minipage}


    \caption{Performance of four methods under different implicit vector dimensions and regularizations.}
    \label{fig:simu_IL}
\end{figure}

To further test the effectiveness of our algorithm, we fix the distribution of $\boldsymbol{U}$ and $\boldsymbol{M}$, as well as the density of the rating matrix $\boldsymbol{R}$.
Let $\lambda \in \{0.05,0.1,0.15\}$, referred to $\mathbf{L1}-\mathbf{L3}$.
Note that $\mathbf{L1}$ corresponds to relatively small penalty terms, and $\mathbf{L3}$ corresponds to relatively large penalty terms.
Let implicit vector dimension $n_f \in \{40,50,60,70\}$, referred to $\mathbf{I1}-\mathbf{I4}$.
We investigate the performance of the algorithm under different regularization parameters and different implicit vector dimensions in Fig. \ref{fig:simu_IL}. The corresponding running times and memory consumption are provided in the Appendix.

As shown in Fig. \ref{fig:simu_IL}, the CORE method still substantially outperforms other sampling methods across all metrics under different implicit vector dimensions and regularizations, and it matches the performance of the full-sample method even at a small subsampling rate.

\subsection{Convergence Performance}
\begin{figure}[h]
    \includegraphics[height=0.4\textheight,width=0.9\textwidth]{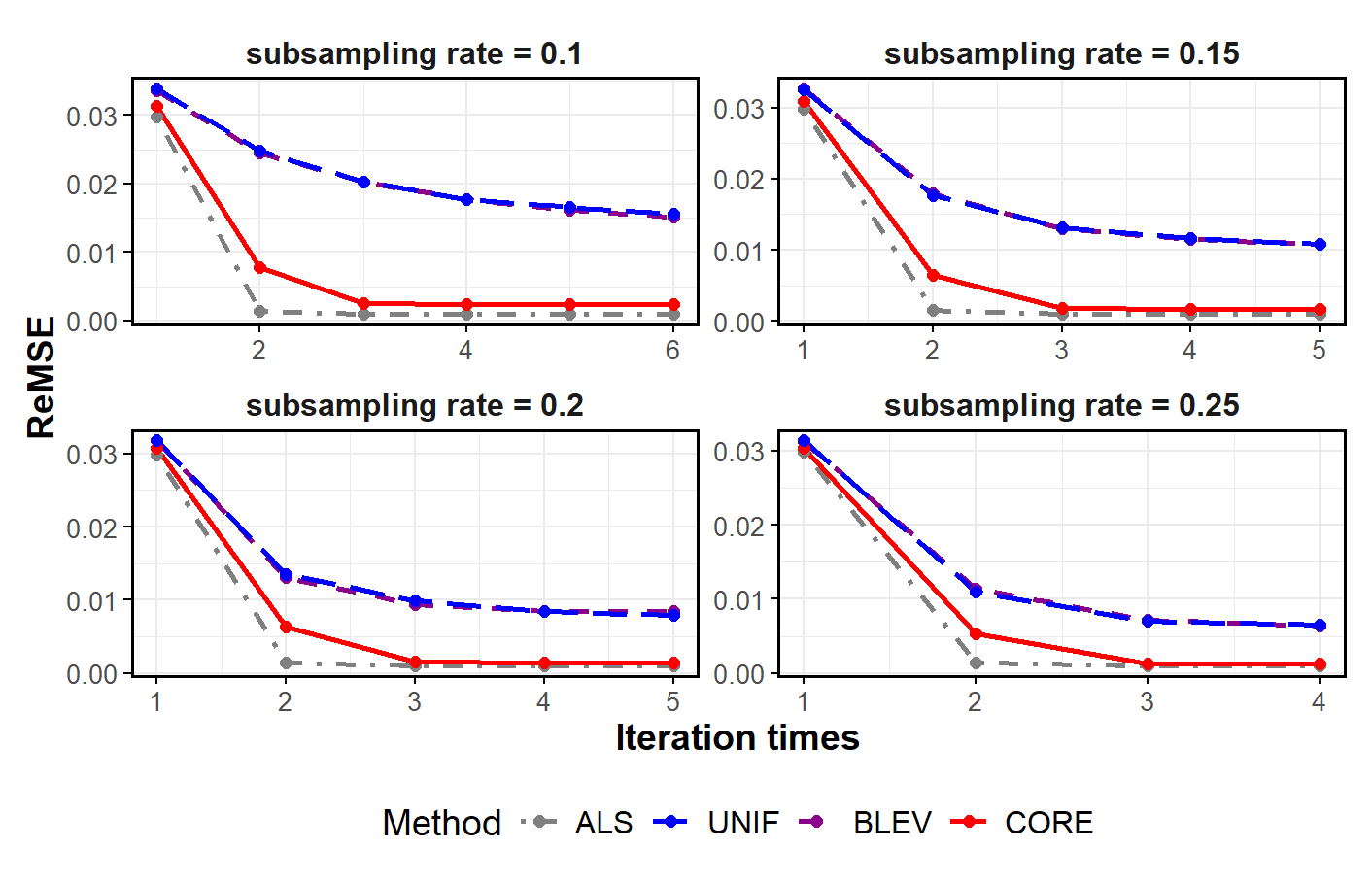}
    \centering
    \caption{ReMSE with respect to the number of iterations.}
    \label{iteration_times_simu}
\end{figure}

In addition to estimation accuracy, we also focus on the convergence speed, as faster convergence requires fewer iterations and thus less computational time. We define convergence as the point where the iteration error falls below $0.01$. Throughout this experiment, the rating matrix dimensionality is kept fixed, the sparsity is set to $20\%$, $\lambda$ is fixed at $0.2$, and the implicit vector dimension is set to $60$. Figure \ref{iteration_times_simu} shows how the ReMSE changes w.r.t the number of iterations for different algorithms under four subsampling rates.

As shown in Fig. \ref{iteration_times_simu}, with the increase of the sampling rate, the convergence speed of all methods accelerates, aligning well with practical expectations. Additionally, we observe that the convergence speed of the CORE method is consistently close to that of the full-sample ALS, and it also requires a similar number of iterations.

\subsection{Computing Time }
Next, we provide the specific runtime of the algorithms. The size of the rating matrix is set to $5000 \times 5000$, with the implicit vector dimension being set to $50$, and the other parameters are consistent with those of $\mathbf{L1, I1}$. Because  BLEV incurs significant overhead in sampling processes, it does not lead to a notable acceleration of ALS. Therefore, we only present the runtime of UNIF here, as its sampling overhead is ignorable. Thus, for the same sampling ratio, it serves as a lower bound for the algorithm's runtime. Results are shown in Table \ref{tb:1}.

\begin{table}[h]
    \centering
    \caption{Computing time for different methods}
    \begin{tabular}{cccccc} 
        \toprule
        Subsampling rate & FULL & 0.1 & 0.15 & 0.2 & 0.25  \\
        \midrule
        CORE & - & 50.50s & 65.82s  & 75.68s & 83.00s \\
        \midrule
        UNIF & - & 36.65s & 50.40s   & 59.63s & 65.14s \\
        \midrule
        ALS & 222.03s & - & -   & - & - \\
        \bottomrule
    \end{tabular}
    \label{tb:1}
\end{table}

As shown in Table \ref{tb:1}, CORE significantly reduces the runtime of ALS at various sampling ratios, with speeds only slightly slower than UNIF. Due to hardware constraints, we were unable to scale the matrix size indefinitely. Additionally, to obtain more stable results, we did not extract additional samples, since we needed to ensure that the number of sampled rows exceeded the latent dimension $n_f$, thus avoiding potential singularities in the calculations. In fact, for very large matrices, the sampling overhead of the CORE method becomes negligible, while the speedup achieved through sparse matrix multiplication becomes increasingly significant.


\section{Real Data Analysis}\label{ch:6}
\subsection{Netflix Competition Data}
The Netflix Competition Data \citep{netflix_dataset} stemmed from a large-scale data mining competition organized by Netflix to identify the most accurate recommender system for predicting user movie ratings. The training data comprises more than 100 million user ratings, contributed by over 480,000 users for 17,700 movies. Each record is stored as a quadruple (user, movie, date, rating), where the rating is an integer from 1 to 5. We employ this dataset to assess the performance of our Core-ALS method.

\subsubsection{Estimation Accuracy}
Given the huge scale of the original dataset and the memory constraints of our computing infrastructure, we constructed a reduced yet representative subset by selecting the top 10,000 users and the top 10,000 movies based on interaction frequency. For each run of our repeated experiments, $80\%$ of the ratings were randomly sampled to form the training set, while the remaining $20\%$ constituted the test set. Subsequently, we treat the items in the test set with ratings greater than or equal to 4 as the true recommendation targets. We evaluated the performance of all compared methods using four commonly adopted metrics:  Relative Mean Square Error (ReMSE),  Prediction ReMSE (P-ReMSE), Hit Ratio at rank 5 (Hit@5), and Normalized Discounted Cumulative Gain at rank 10 (NDCG@10).

\begin{figure}[h]
    \centering

    \begin{minipage}[b]{0.496\textwidth}
        \centering
        \includegraphics[width=\linewidth]{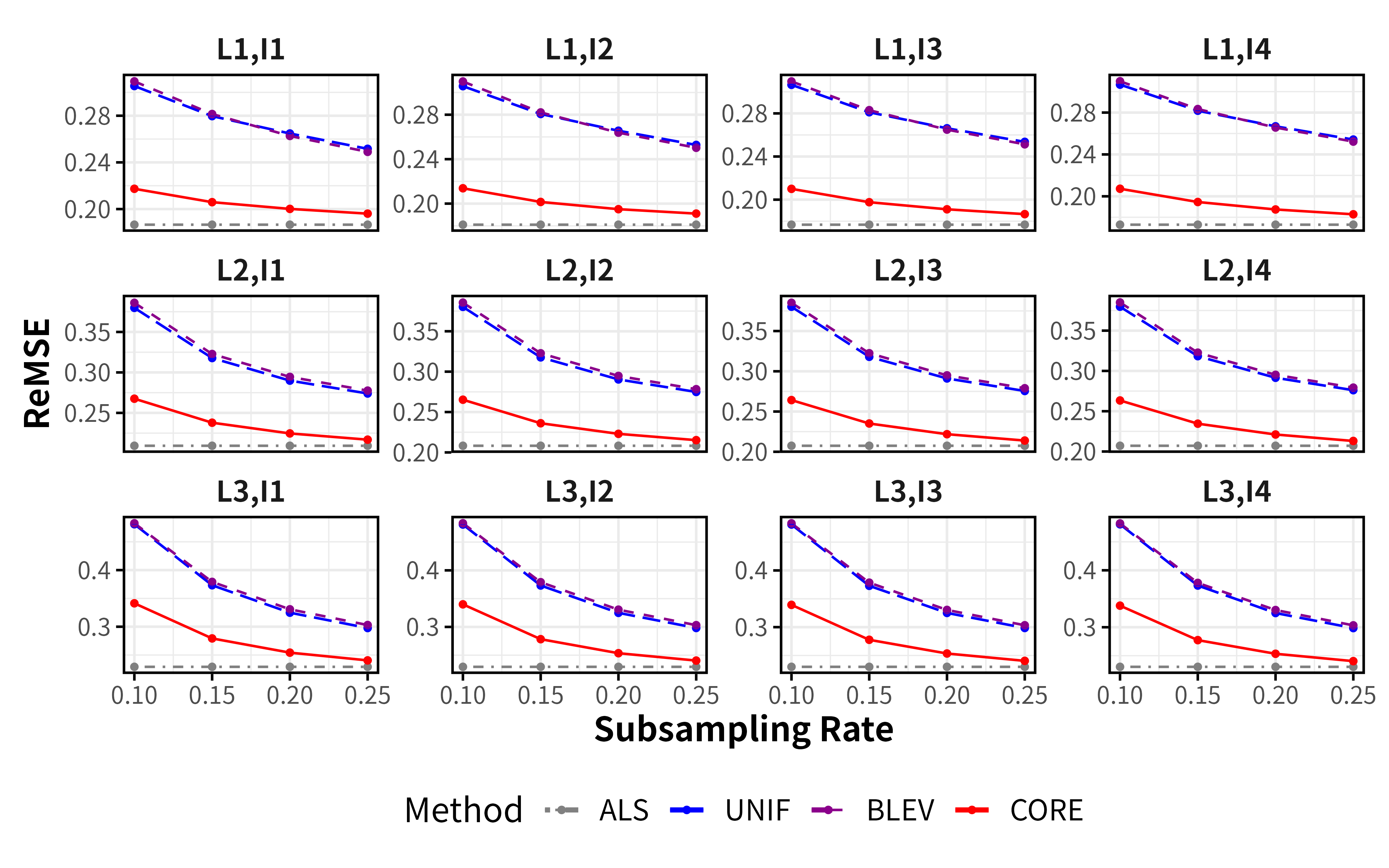}
        \caption*{\footnotesize(a) ReMSE for different L and I.}  
    \end{minipage}
    \begin{minipage}[b]{0.496\textwidth}
        \centering
        \includegraphics[width=\linewidth]{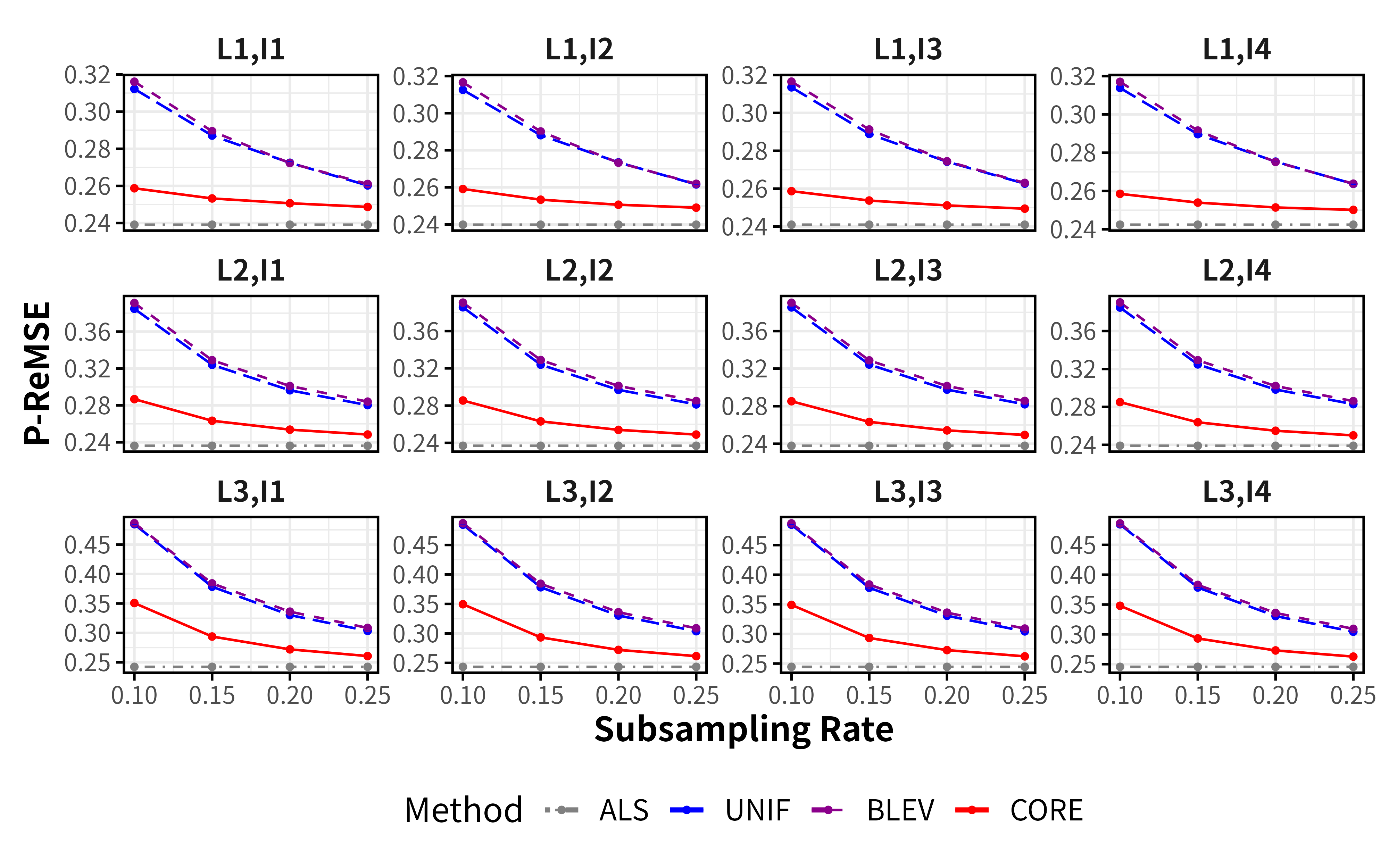}
        \caption*{\footnotesize(b) P-ReMSE for different L and I.}
    \end{minipage}


    \begin{minipage}[b]{0.496\textwidth}
        \centering
        \includegraphics[width=\linewidth]{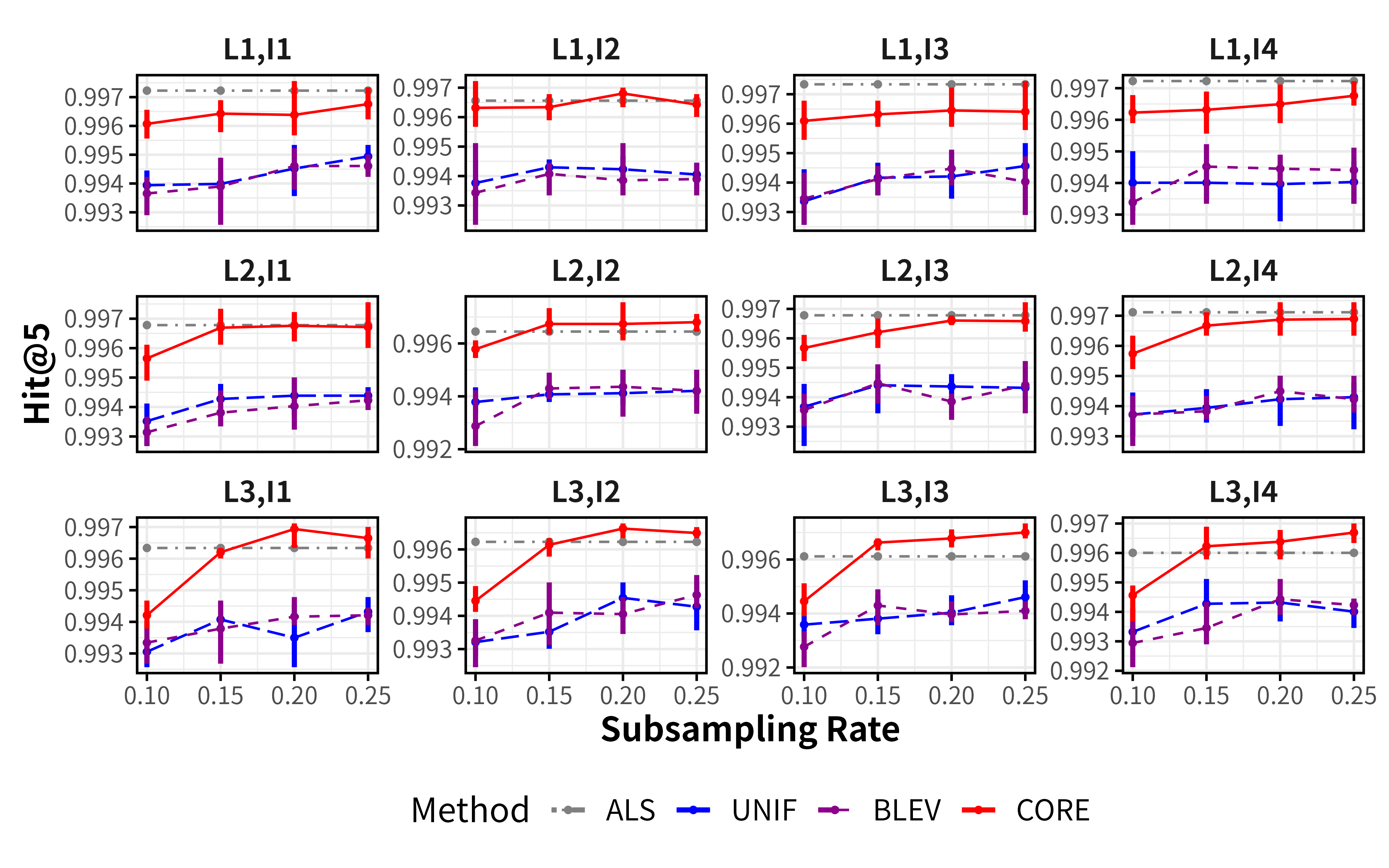}
        \caption*{\footnotesize(c) Hit@5 for different L and I.}
    \end{minipage}
    \begin{minipage}[b]{0.496\textwidth}
        \centering
        \includegraphics[width=\linewidth]{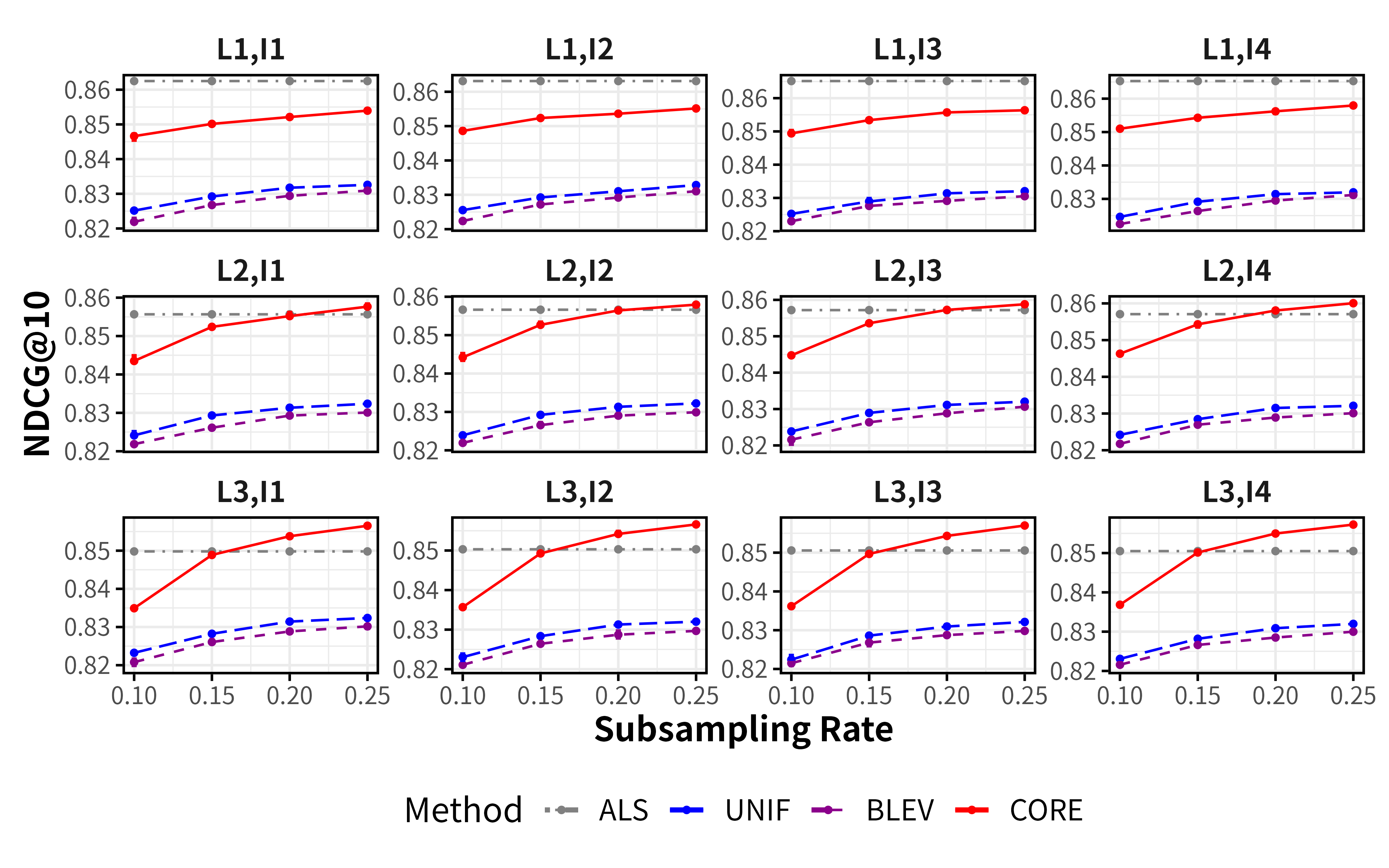}
        \caption*{\footnotesize(d) NDCG@10 for different L and I.}
    \end{minipage}


    \caption{Performance of four methods under Netflix data.}
    \label{fig:netflix_data}
\end{figure}

We also evaluate each method under different subsampling rates, regularization parameters, and implicit vector dimensions. 
We set $\lambda = \{ 0.1, 0.15, 0.2\}$, referred to $\mathbf{L1}-\mathbf{L3}$ and set $n_f = \{20, 25, 30 , 35\}$, referred to $\mathbf{I1}-\mathbf{I4}$.
Figure \ref{fig:netflix_data} shows the relationships between four metrics and the subsampling rate under different regularization parameters and implicit vector dimensions.

As shown in Fig. \ref{fig:netflix_data}, the ReMSE and P-ReMSE of all methods decrease as the subsampling rate increases, which is consistent with expectations. Among all methods, CORE consistently remains closest to the full-sample method. Moreover, as the subsampling rate increases, CORE exhibits an overall upward trend in both Hit and NDCG metrics, demonstrating its ability to achieve recommendation performance comparable to the full-sample method. We have also included additional experiments on the full Netflix dataset in the Appendix to demonstrate that CORE can effectively handle 100 million interactions.

\subsubsection{Computing Time}
For a comparison of runtime, we selected ratings from $40,000$ users for $15,000$ movies. This dataset is highly sparse, with only about $1\%$ of non-missing values. The implicit vector dimension is set to $80$, with the remaining parameters consistent with the settings mentioned earlier.
\begin{table}[h]
    \centering
    \caption{Computing time for different methods}
    \begin{tabular}{cccccc} 
        \toprule
        Subsampling rate & FULL & 0.1 & 0.15 & 0.2 & 0.25  \\
        \midrule
        CORE & - & 137.35s & 162.51s  & 185.72s & 196.95s \\
        \midrule
        UNIF & - & 115.37s & 145.79s   & 155.33s & 168.70s \\
        \midrule
        ALS & 606.15s & - & -   & - & - \\
        \bottomrule
    \end{tabular}
    \label{tb:2}
\end{table}

Although the rating matrix seems large, its extreme sparsity makes its effective size comparable to that of the simulation. Consequently, the results obtained by Table \ref{tb:2} are consistent with those of Table \ref{tb:1}.

\subsection{Bark Texture Image Data}
To illustrate the performance of the Core-ALS method more clearly, we applied it to an image restoration task. We observed that the ALS algorithm demonstrated superior performance in restoring low-rank images. Therefore, we selected several images from the bark texture image dataset \citep{barkvn50_dataset} and randomly occluded a subset of pixels in each image. Subsequently, the masked images were restored using both ALS and CORE algorithms. The restoration results from ALS and CORE were then compared to assess the accuracy and effectiveness of the CORE algorithm in the context of image restoration.

\begin{figure}[h]
    \includegraphics[height=0.55\textheight,width=\textwidth]{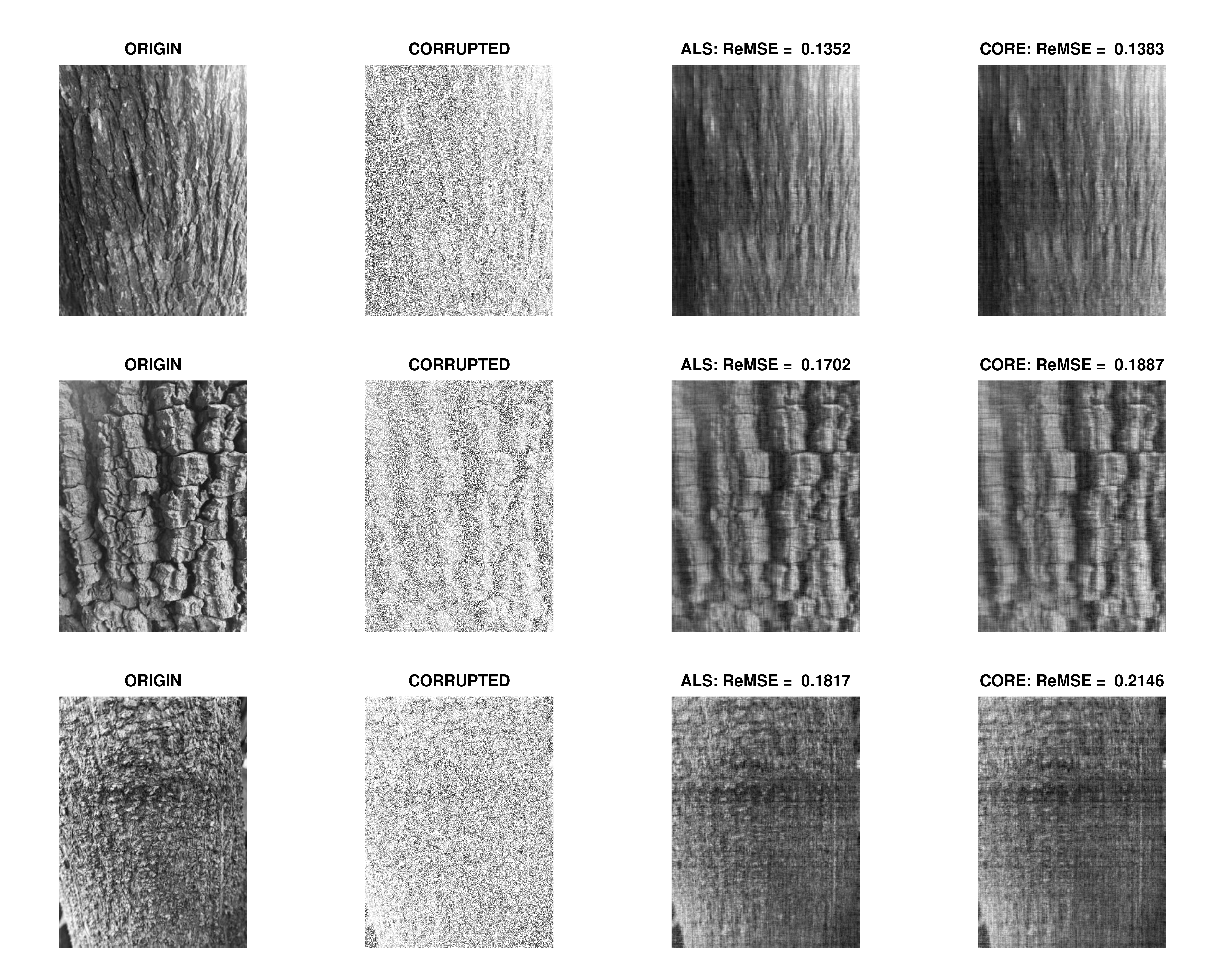}
    \centering
    \caption{Comparison of bark texture image restoration tasks.}
    \label{texture}
\end{figure}

In this experiment, $60\%$ of the pixels in each image were randomly masked. For the ALS algorithm, the dimensionality of the implicit vector was set to 50, with a regularization parameter ($\lambda$) of $0.01$ and $5$ iterations. The CORE algorithm used a sampling proportion of $0.15$, with all other parameters aligned with those of the ALS algorithm. Figure \ref{texture} displays the restoration results for three selected images from the dataset.

As shown in Fig. \ref{texture}, even with a sampling rate as low as $0.15$, the CORE method achieves image recovery performance almost identical to that of ALS.

In conclusion, for the real data, the test results are in good agreement with our simulation results, which reflects the effectiveness of our proposed core-elements algorithm.

\section{Conclusion}\label{ch:7}
This study has introduced an innovative core-elements method for enhancing the efficiency of the alternating least squares (ALS) algorithm used in matrix factorization for recommender systems. The method strategically selects a subset of elements, optimizing computational resources while maintaining the integrity and accuracy of the model predictions.

Our theoretical analyses establish strong guarantees for the approximation and convergence of the proposed method. These results confirm that the core-elements method not only retains the predictive power of the full dataset but does so with significantly reduced computational overhead. This makes the method particularly useful in settings where computational resources are limited or when dealing with extremely large datasets.

Practical applications using both simulated and real-world datasets have demonstrated the effectiveness of the core-elements method. In all tested scenarios, the method performed on par or better than traditional subsampling methods for ALS, especially in terms of computational efficiency. This was evident in our experiments with the Netflix Prize dataset, where the core-elements method consistently showed superior performance compared to other subsampling techniques.

This accelerated algorithm has also prompted several potential avenues for future research. Specifically, determining the optimal hyperparameters that yield the best lower bound for acceleration, analyzing the convergence rates under various sampling probabilities, and exploring richer sampling strategies such as block or row–column hybrid sampling are aspects that remain unexplored in this study. We anticipate that further investigation will enable the development of a robust theoretical framework. Additionally, our current method focuses solely on accelerating ALS in terms of runtime, without improvements in memory usage. Reducing memory overhead is an important direction, and we plan to explore memory-efficient implementations in future work.
Moreover, we aim to extend the accelerated algorithm to low-rank tensor decomposition. \cite{WOS:000458973703072} applied leverage score-based sampling methods to this area, but experimental results indicate that such methods are less effective in matrix factorization, and the computation of leverage scores imposes a considerable computational burden. Therefore, we contend that accelerating core element sampling presents a promising alternative.

Furthermore, with the advancement of computational tools, we are keen to explore the application of tensor decomposition to large-scale, real-world data sets, addressing critical scientific challenges, and furthering our understanding across various research domains.

\section*{Acknowledgments}
This work is supported by Beijing Municipal Natural Science Foundation No. 1232019, National Natural Science Foundation of China Grant No. 12301381, and Renmin University of China research fund program for young scholars.

\section*{Supplementary Material}
\begin{description}
\item[Appendix:] The Appendix contains additional content of the main text and complete proofs of theoretical results, technical details of the theoretical results. Additional numerical experiments and results of various extended versions of Core-ALS are also provided. (appendix.pdf, a pdf file)
\item[Code:] The zip file contains the R code that implements the proposed method and reproduces the numerical results. A README file is included to explain the contents. (code.zip)
\end{description}

\section*{Disclosure Statement}
The authors report there are no competing interests to declare.

{\spacingset{1.2}
\bibliographystyle{chicago}
\bibliography{ref} 
}
\end{document}